\documentclass[a4paper,11pt]{article}%prd,twocolumn

\usepackage{jcappub}

\usepackage{lineno}
\linenumbers

\usepackage{amssymb,amsmath,verbatim,mathtools,needspace,enumitem,etoolbox,graphicx,physics,microtype,afterpage,xspace,tabularx,lmodern,multirow,bm}
\usepackage{dcolumn}
\usepackage{yhmath}

\usepackage{multirow}
\usepackage{float}
\usepackage{booktabs}
\usepackage{gensymb}
\usepackage{lipsum}
\definecolor{TurkishBlue}{HTML}{144893}
\definecolor{romared}{RGB}{142,0,28}

\newcommand{\be}{\begin{equation}}
\newcommand{\ee}{\end{equation}}

\newcommand{\beq}{\begin{eqnarray}}
\newcommand{\eeq}{\end{eqnarray}}

\usepackage{aas_macros}
\usepackage{makecell}
\usepackage{soul}
\usepackage[nolist,nohyperlinks]{acronym}

\usepackage [english]{babel}
\usepackage [autostyle, english = american]{csquotes}
\MakeOuterQuote{"}

\usepackage{color}
\usepackage{tikz}
\usepackage{multirow}
\usepackage{siunitx}
\usepackage[export]{adjustbox}
\usepackage{graphicx}
\usepackage{dcolumn}
\usepackage{bm}
\usepackage[normalem]{ulem}
\usepackage{epstopdf}
\usepackage{tabularx}
\usepackage{suffix}
\usepackage{mathtools}
\usepackage{graphicx}
\usepackage{dcolumn}
\usepackage{bm}
\usepackage[normalem]{ulem}

\graphicspath{ {figures/} }

\usepackage{epstopdf}
\newcommand{\nhat}{\hat{ \mathbf{n}}}

\newcommand{\jhu}{William H.~Miller III Department of Physics and Astronomy, Johns Hopkins University, Baltimore, MD 21218, USA}

\newcommand{\perimeter}{Perimeter Institute for Theoretical Physics, 31 Caroline St N, Waterloo, ON N2L 2Y5, Canada}

\arxivnumber{2312.00118} % Only if you have one
\title{\boldmath Reconstructing patchy helium reionization using the cosmic microwave background and large-scale structure}

\author{Mesut~\c{C}al{\i}\c{s}kan,$^a$}
\author{Neha~Anil~Kumar,$^a$}
\author{Selim~C.~Hotinli,$^{a,b}$}
\author{and~Marc~Kamionkowski$^a$}
\emailAdd{caliskan@jhu.edu}
\emailAdd{nanilku1@jhu.edu}
\emailAdd{shotinli@perimeterinstitute.ca}

\affiliation{$^a$\jhu\\
$^b$\perimeter\\
}

\abstract{The intergalactic helium became fully ionized by the end of cosmic noon ($z\sim2$). 
Similarly to the reionization of hydrogen, helium reionization is expected to be patchy, driven by luminous quasars that ionize the intergalactic gas in their surrounding environment. 
Probing the morphology of ionized electrons during this epoch can provide crucial information about early structure formation, including the clustering and luminosities of quasars, the accretion rates, variability, and lifetimes of active galactic nuclei, as well as the growth and evolution of supermassive black holes.
In this study, we present how measurements of the cosmic microwave background (CMB) can be used to reconstruct the optical-depth fluctuations resulting from patchy helium reionization.
As helium reionization occurred at lower redshifts, upcoming probes of large-scale structure surveys will present a significant opportunity to enhance the prospects of probing this epoch by their combined analysis with the CMB. 
Using a joint information-matrix analysis of hydrogen and helium reionization, we show that near-future galaxy and CMB surveys will have enough statistical power to detect optical-depth fluctuations due to doubly-ionized helium, providing a way of measuring the redshift and duration of helium reionization to high significance.
We also show that modeling uncertainties in helium reionization can impact the measurement precision of parameters characterizing hydrogen reionization.}

\begin{document}
\maketitle
\flushbottom

%%%%%%%%%%%%%%%%%%%%%%
%%%% Intro %%%%%%%%%%%
%%%%%%%%%%%%%%%%%%%%%%

\section{Introduction}

Throughout the cosmic history of structure formation, the intergalactic medium undergoes two significant transitions: the reionization epochs of hydrogen and helium.\footnote{In this work we refer to the reionization of the second electron of helium as ``helium reionization.''}
While the epoch of hydrogen reionization has attracted much attention from the cosmology and astrophysics communities~\citep[see, e.g.,][ for reviews]{Barkana:2000fd,Pritchard:2011xb}, the focus on the reionization of helium has been more limited~\citep[see, however,][]{Wyithe:2002qu,Madau:2015cga,McQuinn:2012bq,Worseck:2014gva,Furlanetto:2007mg,Worseck:2011qk,Sokasian:2001xh,Compostella:2013zya,Oh:2000sg,Furlanetto:2007gn,Furlanetto:2008qy,Dixon:2009xa,LaPlante:2016bzu,Caleb:2019apf,Linder:2020aru,Meiksin:2011bq,Compostella:2014joa,Eide:2020xyi,UptonSanderbeck:2020zla,Bhattacharya:2020rtf,Villasenor:2021ksg,Meiksin:2010rv,Gotberg:2019uhh,LaPlante:2015rea,Syphers:2011uw,Dixon:2013gea,LaPlante:2017xzz,Lau:2020chu,Hotinli:2022jna,Hotinli:2022jnt} despite its observational accessibility. 
Since the second electron in helium has an ionization energy of $\sim 54{\,\rm eV}$, photons expected to be produced by stars do not doubly ionize the intergalactic helium efficiently at cosmological scales. 
Diffuse helium remains singly ionized until populations of active galactic nuclei and quasars start emitting sufficiently energetic photons. 
Consequently, the epoch of helium reionization is expected to take place at lower redshifts ($z\sim3$), a time which is accessible through observations of the ionizing sources.
The morphology of this epoch is closely linked to quasar number counts, clustering and luminosities~\cite{2012ApJ...755..169M,2013ApJ...773...14R,2013ApJ...768..105M,McGreer:2017myu,2022ApJ...928..172P}, the formation of active galactic nuclei, their variability and lifetimes~\citep{Shen:2014rka,Hopkins:2006vv,2017ApJ...847...81S}, as well as early formation of supermassive black holes~\citep{Inayoshi:2019fun}. 
There is now also growing evidence that this epoch will be observationally accessible from joint analyses of the CMB with other cosmological surveys~\citep[e.g.,][]{Hotinli:2022jna,Hotinli:2022jnt}. 

Here, we study the prospects of detecting and characterizing helium reionization by extracting the inhomogeneous reionization signal from joint analyses of upcoming cosmic microwave background (CMB) maps and galaxy surveys. Similar to the epoch of hydrogen reionization, the reionization of helium is anticipated to be anisotropic (or `patchy') due to the formation and growth of ionized bubbles around luminous sources. Unlike the reionization of hydrogen, which is sourced by abundant hot stars, the reionization of helium follows more dispersed sources such as quasars.
This suggests that the morphology of helium reionization (in addition to its redshift) should be different than that of the hydrogen, holding valuable information about the distribution and abundance of quasars (see, for example, Ref.~\citep{Furlanetto:2007mg}). 
The patchiness from both epochs would lead to varying optical depth to reionization across the sky, altering the statistics of the observed CMB maps in multiple ways. These include the screening of the surface of the last scattering, generation of new polarization via Thomson scattering from reionization bubbles, and the kinetic Sunyaev Zel'dovich effect~\citep{Sunyaev1972}. 

In this work, we use the statistical technique developed in Refs.~\citep{Dore:2007bz, 2007ApJ...663L...1H, Dvorkin:2008tf,Dvorkin:2009ah,Feng:2018eal}, which was previously applied to extract the hydrogen reionization signal from the CMB, together with a novel application of cross-correlating the patchy optical depth with 
other surveys of large-scale structure (LSS), such as galaxies.
Given the increasing precision of the upcoming CMB experiments, we fold in the epoch of helium reionization and calculate new detection probabilities, forecasting the future ability to characterize each of the independent epochs.
We construct a quadratic estimator for the modes of the patchy optical-depth field, which separates the patchy reionization signal from CMB in the form of a noisy optical-depth map. 
In our forecasts, we cross-correlate this estimator with tomographic measurements of the galaxy over-density field while also considering its use as a standalone probe.

Unlike the optical depth reconstruction for hydrogen reionization~\citep[e.g.,][]{2007ApJ...663L...1H,Dvorkin:2008tf}, the potential science returns from probing patchy helium reionization optical depth have not yet been extensively studied in the context of CMB.
In this study, we bridge this gap by focusing our forecasts on two upcoming CMB experiments, namely, the stage-4 telescope CMB-S4~\citep{CMB-S4:2016ple, Alvarez:2020gvl,Abazajian:2022nyh} and its stage-5 counterpart CMB-HD~\citep{Sehgal:2019ewc, CMB-HD:2022bsz,MacInnis:2023vif}. These upcoming experiments, alongside galaxy surveys such as Vera Rubin Observatory (LSST)~\citep{LSST:2008ijt,LSSTDarkEnergyScience:2018jkl} and MegaMapper~\citep{Schlegel:2019eqc},  stand to transform our understanding of cosmology with an influx of high-precision data in the upcoming years. Among the new windows of opportunity that are being opened by these experiments, the prospects of using the CMB as a cosmological `back-light' to probe LSS from signals sourced by interactions between CMB photons and the intervening cosmological structures particularly motivates this work.

These programs include the reconstruction of the lensing potential
\citep[see, e.g.,][for a review]{Lewis:2006fu}, radial and transverse velocity
\citep[][]{Cayuso:2021ljq,Smith:2018bpn,Hotinli:2018yyc,AnilKumar:2022flx,Kumar:2022bly,Hotinli:2021hih,Hotinli:2020csk,Munchmeyer:2018eey,Deutsch:2017ybc,Hotinli:2023ywh}, and quadrupole fields~\citep[][]{Deutsch:2017ybc,Hotinli:2022wbk,Lee:2022udm,Namikawa:2023zux,Deutsch:2017cja},  as well as the patchy optical depth, which we focus on in this paper. 
While previous studies have explored the prospects of joint analyses between reconstructed optical depth and tracers of LSS, such as 21-cm intensity mapping and Compton-$y$ maps, in the context of hydrogen reionization and for probing the circumgalactic medium~\citep{Meerburg:2013dua,Roy:2019qsl,Orlando:2023dgt,Roy:2022muv}, these methods have not yet been applied to probing helium reionization.
Given that helium reionization takes place at lower redshifts---a range well-covered by upcoming galaxy surveys~\citep[][]{LSST:2008ijt,Sailer:2021yzm}, which will detect high numbers of galaxies---it stands to benefit significantly from the statistical power provided by cross-correlations with the galaxy distribution, in contrast to hydrogen reionization.

Little is known about the precise characteristics of helium reionization, such as its morphology, timing, and duration. While surveys measuring helium and hydrogen Lyman-$\alpha$ forests can, in principle, help characterize this epoch, the analyses are subject to a range of systematic and astrophysical uncertainties, such as those affecting the inferred flux levels of the Lyman-$\alpha$ forests~\citep{2011MNRAS.410.1096B,2012AJ....143..100S,Boera:2014sia,Telikova:2019xph}, for example. 

Specifically, measurements of hydrogen Lyman-$\alpha$ serve only as an indirect probe of helium reionization, inferring insights through the thermal history of the intergalactic medium in a model-dependent manner, and are subject to systematic uncertainties.
On the other hand, these measurements are also significantly obscured by intervening Lyman-limit systems at lower redshifts.
Although further simulations and analyses will be needed to understand whether these issues impose a significant limit on the prospects of probing helium reionization from Lyman-$\alpha$ forests, it is nevertheless suggestive that additional probes will be valuable for the unambiguous characterization of this epoch.

In addition to enhancing our understanding of the astrophysics of quasars and active galactic nuclei, probing helium reionization offers additional benefits for cosmological inference. 
For example, the determination of the total change in the free-electron fraction during this epoch serves as an indicator of the primordial helium abundance $Y_p$. In turn,
improved measurements of $Y_p$ can significantly enhance our understanding of weak interaction rates, neutron lifetime, and Big Bang nucleosynthesis, as discussed in Refs.~\citep[][]{Pitrou:2018cgg,Yeh:2022heq}. 
Such advancements provide valuable insights into the intricate details of our cosmological history.
As the effective number of degrees of freedom of light relic species $N_{\rm eff}$ and $Y_p$ affect the small-scale CMB damping tail in similar ways~\citep{PhysRevD.87.083008}, improving the $Y_p$ measurement also improves measurements of $N_{\rm eff}$, a driving science goal of the upcoming CMB experiments.\footnote{This improvement is achieved, for instance, by breaking the degeneracy between $Y_p$ and $N_{\rm eff}$~\citep[][]{Green:2016cjr,Hotinli:2021umk,Lee:2023uxu}.} 

Finally, here we model electrons in the ionized media as tracers of dark matter. Recent observations~\citep[e.g.,][and references therein]{2021PhRvD.103f3513S,ACT:2024rue}, however, suggest that electrons do not necessarily follow dark matter on sub-halo scales.
This has implications for the detectability of helium reionization if these uncertainties need to be marginalized over.
Nevertheless, upcoming measurements of Sunyaev Zel'dovich effects from joint analyses of CMB and galaxy surveys stand to provide high-precision measurements of electron density profiles and electron-galaxy correlations, likely reducing these uncertainties. 
Our forecasts are presented in such a way that the effects of these uncertainties on our central results can be inferred.

This paper is organized as follows. 
In Sec.~\ref{sec:He_reionization_model}, we introduce our model of patchy reionization characterizing the redshift evolution of the ionized regions during both helium and hydrogen reionization. 
We introduce the primary model of the integrated, auto-correlation signal $C_{\ell}^{\tau\tau}$, accounting for optical-depth fluctuations sourced by both epochs. 
Furthermore, we also present the central model for the cross-correlation of the patchy-reionization optical-depth field with large-scale, tomographic galaxy survey data.
We conclude this modeling section by deriving the modifications to CMB spectra arising from patchy reionization. 
We present the estimator for the patchy optical-depth reconstruction in Sec.~\ref{sec:tau_estimator}.
With the signal and noise established, we finally show our results from the forecasts in Sec.~\ref{sec:forecasts}.
We introduce the assumed experiment specifications and the resulting measurement noise for near-future CMB experiments and galaxy surveys. 
We then forecast the ability of these future surveys to probe optical-depth fluctuations sourced by helium reionization via computation of the measurement SNR. 
We also forecast the measurability of the morphology of this epoch by presenting results from our information-matrix campaign. 
These results include the anticipated fractional errors on the model parameters and an exploration of how our chosen fiducial parameters impact the measurability of both helium and hydrogen reionization.
We conclude with a discussion in Sec.~\ref{sec:discussion}. Appendix~\ref{sec:binned_optical_depth} contains more details on the derivation of the binned auto-power spectrum of the optical-depth field, as well as its cross-correlation with the galaxy density field. Appendix~\ref{sec:estimator_detail} contains supplementary information on the derivation of the noise in the CMB-reconstructed optical-depth field. Finally, Appendix~\ref{sec:imporvements_beyond_ilc} contains selected forecast results for experiment configurations assuming a more optimistic foreground cleaning methodology.

\section{Helium reionization}\label{sec:He_reionization_model}

In this section, we detail the modeling of the inhomogeneous optical depth, calculate the pertinent angular power spectra, and examine the observable impacts of patchy reionization on the CMB.
Section~\ref{sec:model_optical_depth} starts with a definition of the optical depth of photons scattering off free electrons originating from hydrogen and helium reionization. 
We then model the mean ionization fraction and describe the assumed size distribution of ionized regions. 
Section~\ref{sec:var_params} explores the influence of varying helium reionization parameters. 
In Sec.~\ref{sec:Cl_tautau}, we compute the angular power spectrum of the average optical depth for both hydrogen and helium, incorporating the cross-power term between the two. 
Section~\ref{sec:Cl_taug} is dedicated to deriving the angular cross-power between galaxy and ionized electron fluctuations and presents a comparative analysis of all the binned power spectra. 
The section concludes with a comprehensive discussion of the effects of patchy reionization on the CMB in Section~\ref{sec:effects_on_CMB}.

\subsection{Modeling the inhomogeneous optical depth}\label{sec:model_optical_depth}

The optical depth of photons scattering off of free electrons sourced by reionization, out to redshift $z$, is

\be
\tau_\theta(\nhat, z)=\sigma_T n_{\theta,0}\int_0^z\frac{\dd z'(1+z')^2}{H(z')}x_e^\theta(\nhat,z')\,,
\ee

where we use the symbol $\theta\in\{\rm H, He\}$ to distinguish between the contributions from reionization of hydrogen and helium, respectively. 
Here, $\sigma_T$ is the Thomson scattering cross section, $H(z)$ is the Hubble parameter at $z$, $x_e(\nhat,z)$ is the ionization fraction at redshift $z$ in the line-of-sight direction $\nhat$. 
We take $n_{\theta,0}=\{n_{p,0},f_{\rm He}n_{p,0}\}$, where $n_{p,0}$ is the number density of protons today, and $f_{\rm He}$ is the fraction of helium to hydrogen atoms satisfying $f_{\rm He}\simeq0.08$.

For each of the epochs, we model the mean ionization fraction with a hyperbolic tangent
\be \label{eq:ion_fraction}
\bar{x}^{\theta}_e(z)=\frac{1}{2}\left[1-\tanh\left(\frac{y(z)-y_{\rm re}^{\theta}}{\Delta_y^{\theta}}\right)\right]\,,
\ee
where $y(z)=(1+z)^{3/2}$, and $y_{\rm re}$ and $\Delta_y$ are free parameters of the reionization model which roughly specify the time and duration of the reionization process.
Although simple, this functional form has been used by various studies focusing on patchy reionization~\citep[e.g.,][]{2008PhRvD..78b3002L,Planck:2016mks,PhysRevD.95.023513} and enables us to explore the impact of various reionization times and durations (i.e., when the epoch at which the reionization starts and how fast the media transitions from nonionized to fully reionized) on the detection signal-to-noise ratio (SNR) and parameter estimation, as we explain in later sections.

For hydrogen reionization, we choose the fiducial values $\{y_{\rm re}^{\rm H},\Delta^{\rm H}_y\}=\{27.0,7.0\}$, corresponding to a central redshift of $z\sim 8$ and a duration spanning $11 \gtrsim z \gtrsim 5$. 
This fiducial choice is not only physically motivated by the approximate redshifts of star-formation and measurements of the Lyman-$\alpha$ forests \cite{Barkana:2000fd,Pritchard:2011xb}, but is also consistent with recent CMB measurements of the optical depth to recombination setting $\bar{\tau}\approx 0.060$ \citep{Planck:2015fie, Planck:2018vyg}.
Similarly, for the helium counterpart, we choose $\{y_{\rm re}^{\rm He},\Delta^{\rm He}_y\} = \{8.0,3.1\}$, corresponding to a central redshift of $z\sim 3$ and a duration spanning $5 \gtrsim z \gtrsim 1$. 
This choice is primarily motivated by the expected redshifts of galaxy and quasar formation and its impact on helium reionization \citep{2006PhR...433..181F}. Note that the relatively low abundance of helium allows more freedom in the characterization of $\bar{x}_e^{\rm He}$, with minimal impact on the average optical depth to reionization.

\begin{figure}[h!]
    \centering
    \includegraphics[width=0.7\textwidth]{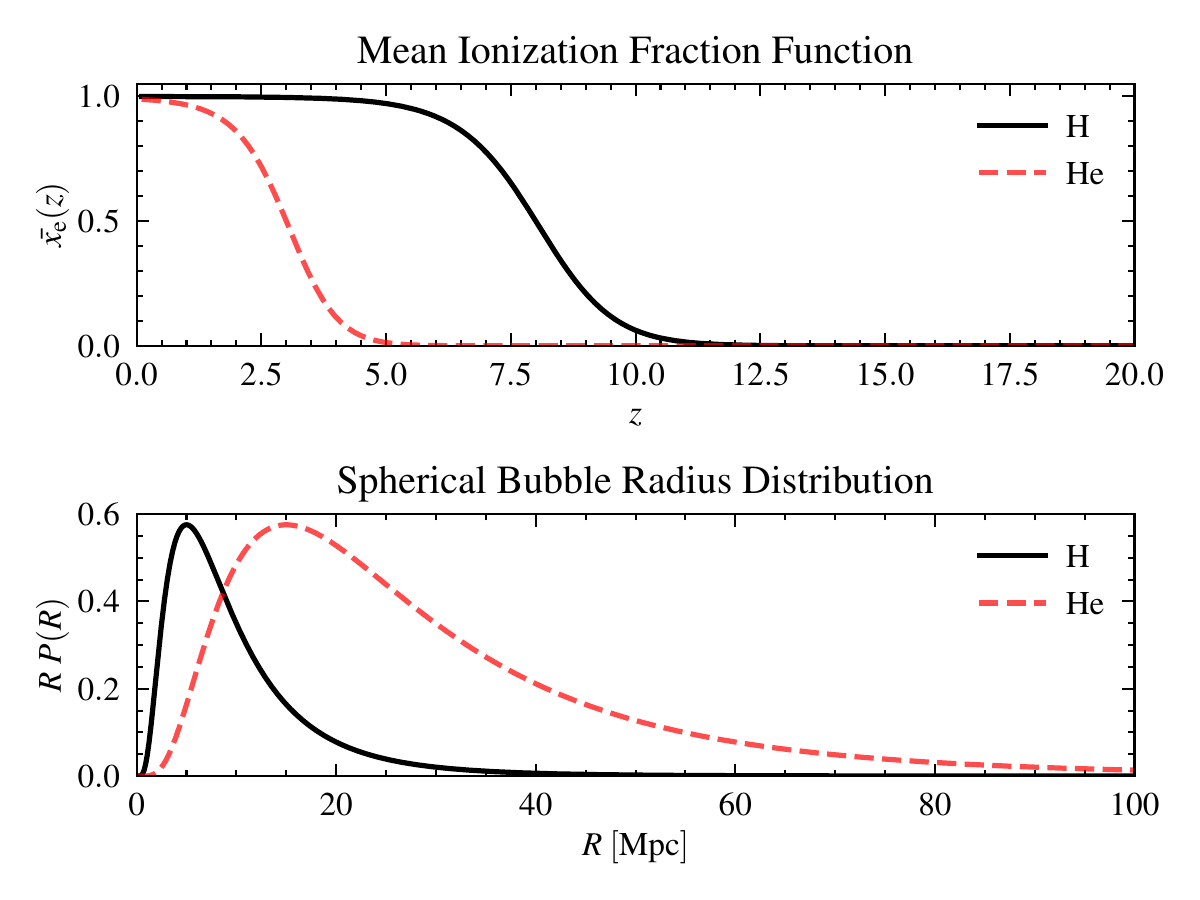} 
    \caption{\textit{Upper panel:} mean ionization fraction $\bar x_e (z)$ for hydrogen (H) and helium (He) as a function of redshift $z$. 
    \textit{Lower panel:} spherical bubble radius distribution $R\, P(R)$ as a function of bubble radius $R$.
    The \textit{patchy reionization} model is based on Eqs.~\eqref{eq:ion_fraction}~and~\eqref{eq:bubble_radius_distb}, and the fiducial model parameters describing the ionization fraction functions and the bubble radius distribution 
    are listed in Tab.~\ref{tab:set_of_params} for each of the epochs.}
    \label{fig:ion_fraction}
\end{figure}

Consistent with existing analysis and modeling of hydrogen reionization~\citep[e.g.,][]{2004ApJ...613....1F,Mortonson:2006re,2007ApJ...654...12Z,2007ApJ...669..663M,Dvorkin:2008tf,Bianchini:2022wte}, we represent ionized regions as spherical `bubble'-like volumes centered at the ionizing source. 
To allow for bubbles of various sizes, we assume that the bubbles have radius $R$, distributed according to a log-normal distribution as follows:
\be \label{eq:bubble_radius_distb}
P(R)=\frac{1}{R}\frac{1}{\sqrt{2\pi\sigma_{\ln R}^2}}e^{-[\ln(R /\bar{R})]^2/(2\sigma_{\ln{R}}^2)}\,,
\ee
where $\bar{R}$ is the characteristic size of the bubbles, and $\sigma_{\ln{R}}$ is the width of the distribution. 
In addition to this, we assume that the number density of reionization bubbles is a biased tracer of matter fluctuations on large scales, and thus we model the spatial distribution of bubbles with a bubble bias $b^\theta$.
Following Ref.~\citep{Dvorkin:2008tf}, we set $\{\bar{R}^{\rm H},\sigma^{\rm H}_{\ln R}\}=\{5\ \rm{Mpc},\, \ln(2)\}$ for hydrogen reionization.
On the other hand, since helium reionization is sourced by luminous quasars and active galactic nuclei emitting hard photons, the typical size of bubbles during the reionization of helium is anticipated to be larger than bubbles forming during the reionization of hydrogen~\citep[e.g.,][]{Furlanetto:2007mg}.
Therefore, we choose $\{\bar{R}^{\rm He},\sigma^{\rm He}_{\ln R}\}=\{15\ \rm{Mpc},\, \ln(2)\}$ for the helium reionization.

The upper panel of Fig.~\ref{fig:ion_fraction} shows the mean ionization fraction as a function of the redshift $z$ for hydrogen (helium) based on these fiducial parameters in solid black (dashed red) line, while the lower panel represents the spherical bubble radius distribution $R P(R)$ as a function of radius $R$.
As seen, helium reionization starts much later ($z \sim 5$) compared to that of hydrogen and goes through a period of maximal patchiness around $z\sim 3$.
As the lower panel indicates, there is a lower probability for helium bubbles to be small ($R \lesssim 5\ \rm{Mpc}$). 
Although most bubbles are with $R \sim 15\ \rm{Mpc}$, the helium bubbles can be much larger than hydrogen bubbles (up to $R \lesssim 100\ \rm{Mpc}$).

\subsection{Varying the model parameters}\label{sec:var_params}

Our choice of fiducial helium reionization parameters $\{y_{\rm re}^{\rm He}, \Delta_y^{\rm He}, \bar R^{\rm He}, \sigma_{\ln R}^{\rm He}\} \\ = \{8.0, 3.1, 15.0\ {\rm{Mpc}}, \ln{(2.0)} \}$ is informed by the expected time and duration of helium reionization, as well as the impact of the difference in ionizing sources on the bubble size distribution~\citep[e.g.,][]{Furlanetto:2007mg}.
Nonetheless, there is great uncertainty as to what these parameters actually are~\citep{2011MNRAS.410.1096B,2012AJ....143..100S,Boera:2014sia,Telikova:2019xph}.
The simplicity of our modeling enables us to explore a wide range of possible reionization configurations by scanning the parameter space of helium reionization parameters in the relevant ranges.

\begin{figure}[h!]
    \centering
    \includegraphics[width=1.0\textwidth]{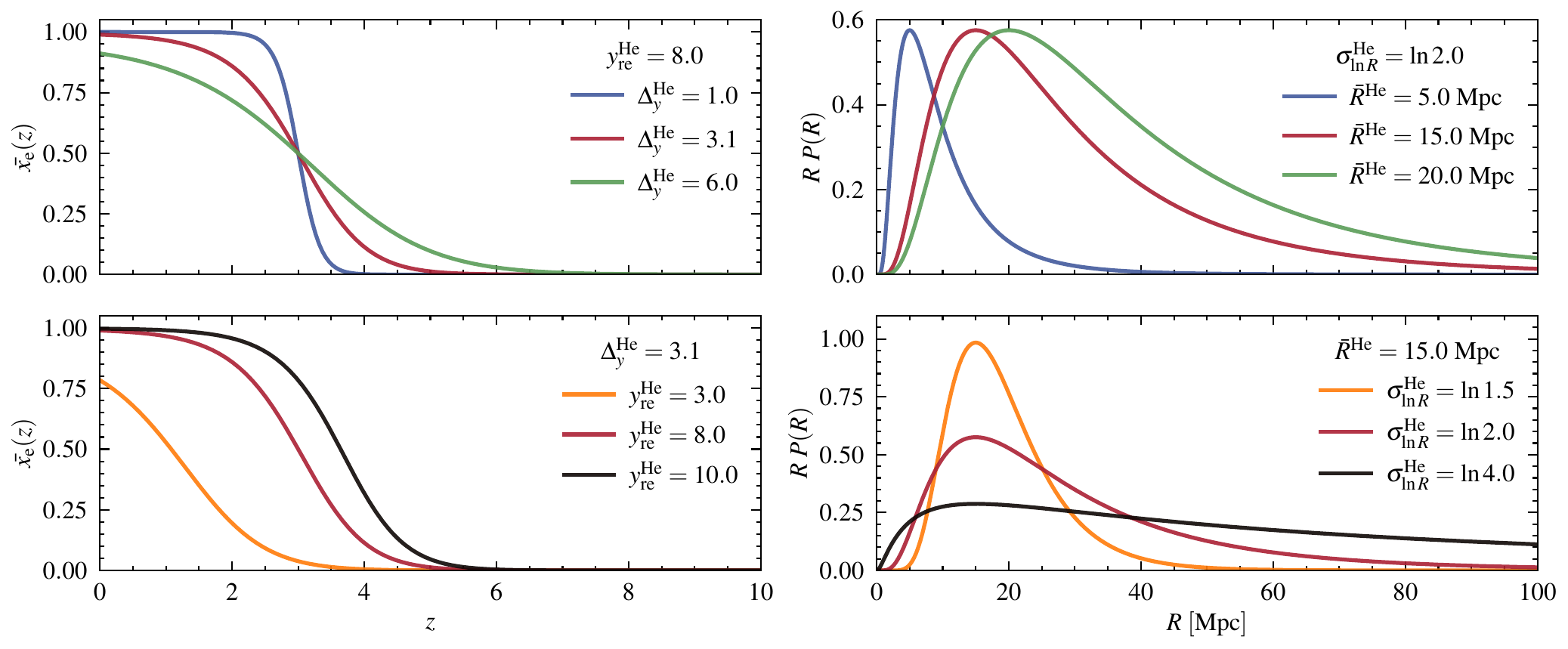} 
    \caption{\textit{Left panels:} average ionization fraction $\bar x_e (z)$ as a function of redshift $z$ for various model parameters characterizing the patchy helium (He) reionization. 
    In the upper panel, $y_{\rm re}^{\rm He} = 8.0$ while $\Delta_y^{\rm He}$ is varied.
    In the lower panel, the former is varied while the latter is fixed at $\Delta_y^{\rm He} = 3.1$.
    \textit{Right panels:} spherical bubble radius distribution $R\, P(R)$ as a function of bubble radius $R$ for various model parameters. Similar to the left panels, $\sigma_{\ln R}^{\rm He}$ ($\bar R^{\rm He}$) is fixed in the upper (lower) panel while the other parameter is varied.
    In each case, the red line shows the results for the set of fiducial parameters.
}
    \label{fig:He_variation}
\end{figure}

The left panels in Fig.~\ref{fig:He_variation} show the average ionization fraction $\bar x^{\rm He}_e (z)$ for various model parameters. 
In the upper left panel, $y_{\rm re}^{\rm He}$ is fixed at the fiducial value of $8.0$ while $\Delta_y^{\rm He}$ is varied between $1.0$ and $6.0$ in blue and green, respectively.
On the other hand, in the bottom left panel, $\Delta_y^{\rm He}$ is fixed at the fiducial value of $3.1$ while $y_{\rm re}^{\rm He}$ is varied between $3.0$ and $10.0$ in orange and black, respectively. 
Given the two plots on the left of Fig.~\ref{fig:He_variation}, we can understand the effects of changing $\Delta_{y}^{\rm He}$ and $y_{\rm re}^{\rm He}$ on the redshift evolution of $\bar{x}^{\rm He}_e(z)$.
The upper left panel indicates that the parameter $\Delta_y^{\rm He}$ determines how smoothly the transition to reionization occurs. 
In other words, the effect of lowering (increasing) $\Delta_y^{\rm He}$ is roughly speeding up (slowing down) the rate at which helium ionizes as a function of redshift. 
Note that changing $\Delta_y^{\rm He}$, with $y_{\rm re}^{\rm He}$ fixed, does not affect the mean reionization redshift [redshift at which $\bar{x}_e^{\rm He}(z) \approx 0.5$].
On the contrary, the bottom left panel suggests that the parameter $y_{\rm re}^{\rm He}$ alters the midpoint's location, essentially shifting the mean redshift of reionization.
Increasing (decreasing) $y_{\rm re}^{\rm He}$ shifts the entire function to the right (left) in redshift, thus changing the time at which reionization occurs.
These minimum and maximum ranges are determined to account for earlier/ later helium reionization epochs as well as faster/ slower transitions to fully reionized helium, and represent the limits of the ranges we consider in our forecasts.

The right panels in Fig.~\ref{fig:He_variation} show the spherical bubble radius distribution for various parameter values.
In the upper right panel, the width of the distribution is fixed, while the characteristic bubble size $\bar R$ is varied between $5.0\ \rm{Mpc}$ and $20.0\ \rm{Mpc}$ in blue and green, respectively.
In the lower right panel, the characteristic size is set to the fiducial value while the width of the distribution is changed between $\ln{1.5}$ and $\ln{4.0}$ in orange and black, respectively. 
These limits are chosen to account for much larger (smaller) helium reionization bubbles than expected, account for uncertainty in the width of the distribution, and represent the limits of the ranges we consider in our forecasts.
In each of the subplots, the red line represents the results for the set of fiducial parameters previously outlined for ease of comparison.

\subsection{Angular power spectrum of the optical depth}\label{sec:Cl_tautau}

Given that the observed optical depth $\tau$ will get contributions from both the epochs of reionization, we use the derivations in Refs.~\citep{Dvorkin:2008tf, Mortonson:2006re} to write the angular power spectrum of the optical depth as a line-of-sight integral as follows: 
\begin{align}
\label{eq:cl_tautauall}
C_\ell^{\tau\tau} = \underbrace{\int_0^{\chi_*}\dd\chi\frac{\sigma_T^2n_{p,0}^2}{a^4\chi^2}P^{\rm H}_{\Delta_e\Delta_e}(\chi,k)}_{\mathrm{H-only}}
&+ \underbrace{\int_0^{\chi_*}\dd\chi\, f_{\rm He}^2 \frac{\sigma_T^2 n_{p,0}^2}{a^4\chi^2}P^{\rm He}_{\Delta_e\Delta_e}(\chi,k)}_{\mathrm{He-only}} \nonumber \\
&+ \underbrace{2\int_0^{\chi_*}\dd\chi\, f_{\rm He} \frac{\sigma_T^2n_{p,0}^2}{a^4\chi^2}\bar{x}_e^{\rm He}\bar{x}_e^{\rm H}P(\chi,k)}_{\mathrm{H-He\ cross}}\,.
\end{align}
where $a$ is the cosmic scale factor, $\chi_*$ is the comoving distance to recombination, $k$ is the Fourier wave number, $P(\chi, k)$ is the nonlinear matter power spectrum, and we use the Limber approximation~\citep{1953ApJ...117..134L}, setting $k=\ell/\chi$. Here $P_{\Delta_e\Delta_e}^{\rm H}$ and $P_{\Delta_e\Delta_e}^{\rm He}$ are the power spectra of hydrogen and helium ionization fluctuations, respectively, which we define next.
Note that we arrive at this model by categorizing the free-electrons sourced by helium reionization separately from those sourced by hydrogen, with the third term in the above equation accounting for correlations in the distributions of the two separate `populations.'
Therefore, the helium-fraction factor appropriately weights each of the helium-dependent terms, in order to account for its relatively low abundance.

\begin{table}[t!]
    \centering
    \caption{The set of reionization model parameters for hydrogen and helium and the galaxy bias factor. The middle columns present the set of fiducial values while the right-side columns show the range of values used in the forecasts described in Sec.~\ref{sec:forecasts}.}
    \label{tab:set_of_params}
    \vspace{10pt}
    \begin{minipage}{\textwidth} 
    \renewcommand{\arraystretch}{1.2} 
    \centering
    \small
    \begin{tabular}{c|c c|c c}
    \hline \hline
    Parameter & \multicolumn{2}{c|}{Fiducial Value} & \multicolumn{2}{c}{Range of Values} \\ \cline{2-5}
              & \ Hydrogen \ & Helium \ & \ Hydrogen \ &  Helium \ \\ \hline
    $y_{\rm re}^{\theta}$                      & $27.0$        & $8.0$          & \texttt{Fixed}        & $3.0 \leq y_{\rm re}^{\rm He} \leq 10.0$      \\
    $\Delta_y^{\theta}$                        & $7.0$         & $3.1$          & \texttt{Fixed}        & $1.0 \leq \Delta_y^{\rm He} \leq 6.0$      \\
    $\bar{R}^{\theta}$                         & $5\ \rm{Mpc}$ & $15\ \rm{Mpc} \ \ $ & \texttt{Fixed}   & $5.0\ \textrm{Mpc} \leq \bar{R}^{\rm He} \leq 20.0\ \textrm{Mpc}$      \\
    $\sigma_{\ln{R}}^{\theta}$                 & $\ln 2.0$     & $\ln 2.0$      & \texttt{Fixed}        & $\ln{1.5} \leq \sigma^{\rm He}_{\ln{R}} \leq \ln{4.0}$      \\
    Bubble bias $b^{\theta}$          & 6.0           & 6.0            & \texttt{Fixed}        & \texttt{Fixed}      \\
    Galaxy bias factor $b_{g,0}$      & \multicolumn{2}{c|}{0.95}      & \multicolumn{2}{c}{\texttt{Fixed}} \\ \hline \hline
    \end{tabular}
    \end{minipage}
\end{table}

To model the power-spectra $P_{\Delta_e\Delta_e}^\theta$ of ionization fluctuations, we assume that the basic morphology of reionization remains the same across the two epochs, and therefore, the functional form for the spectra are identical: 
\be
\begin{split}
P_{\Delta_e\Delta_e}^{\theta}(\chi,k)&=P^{1b,\, \theta}_{\Delta_e\Delta_e}(\chi,k)+P^{2b,\, \theta}_{\Delta_e\Delta_e}(\chi,k)\,,\\
\end{split}
\ee
where, the 2-bubble $P^{2b,\, \theta}_{\Delta_e\Delta_e}(\chi,k)$ and 1-bubble $P^{1b,\, \theta}_{\Delta_e\Delta_e}(\chi,k)$ contributions are
\be
\begin{split}
P^{2b,\, \theta}_{\Delta_e\Delta_e}(\chi,k)&=[(1-\bar{x}_e^{\theta})\ln(1-\bar{x}_e^{\theta}){{b^\theta}}\,I^{\theta}(k)-\bar{x}_e^{\theta}]^2 P(\chi,k)\,,\\
P^{1b,\, \theta}_{\Delta_e\Delta_e}(\chi,k)&=\bar{x}_e^{\theta}(1-\bar{x}_e^{\theta})[F^{\theta}(k)+G^{\theta}(k)]\,.
\end{split}
\ee
Here, the average ionization fraction $\bar{x}_e^{\theta}(z)$ varies from zero to unity according to Eq.~\ref{eq:ion_fraction} (cf.~Fig.~\ref{fig:ion_fraction}), 
$b^\theta$ is the bubble bias and is set to $6.0$ for both hydrogen and helium as a simplifying assumption. It is crucial to acknowledge that in reality, the bubble bias for helium might deviate from that of hydrogen, which could impact the measurability of the model parameters. Specifically, a higher (lower) helium bubble bias value could be expected to yield enhanced (reduced) results.
The terms $\bar{x}_e^{\theta}$, $I^{\theta}(k)$, $G^{\theta}(k)$, $F^{\theta}(k)$ each depend on the details of the reionization: the time and duration $\{y_{\rm re}^{\theta},\Delta_y^{\theta}\}$ as well as the bubble size distribution $\{\bar{R}^{\theta},\sigma^{\theta}_{\ln R}\}$. These spectra are modeled following Ref.~\citep{Mortonson:2006re}.
%\footnote{The final term in Eq.~\eqref{eq:cl_tautauall} comes from the free-electron density fluctuations which we assume follow matter fluctuations, often referred to as the `late-time' contribution to the electron fluctuations. The power-spectrum of the free-electron fluctuations at the limit $z\rightarrow0$ satisfies $P_{ee}(\chi,k)=(1+f_{\rm He})^2n_{p,0}^2P(k)$.} 
The set of fiducial parameters used in modeling the angular power spectrum and the respective range of values used in the forecasts of Sec.~\ref{sec:forecasts} can be seen in Tab.~\ref{tab:set_of_params}. In combination, there are 11 parameters (5 for the description of helium reionization, 5 for hydrogen, and the galaxy bias factor $b_{g,0}$).

\begin{figure}[h!]
    \centering
    \includegraphics[width=0.5\textwidth]{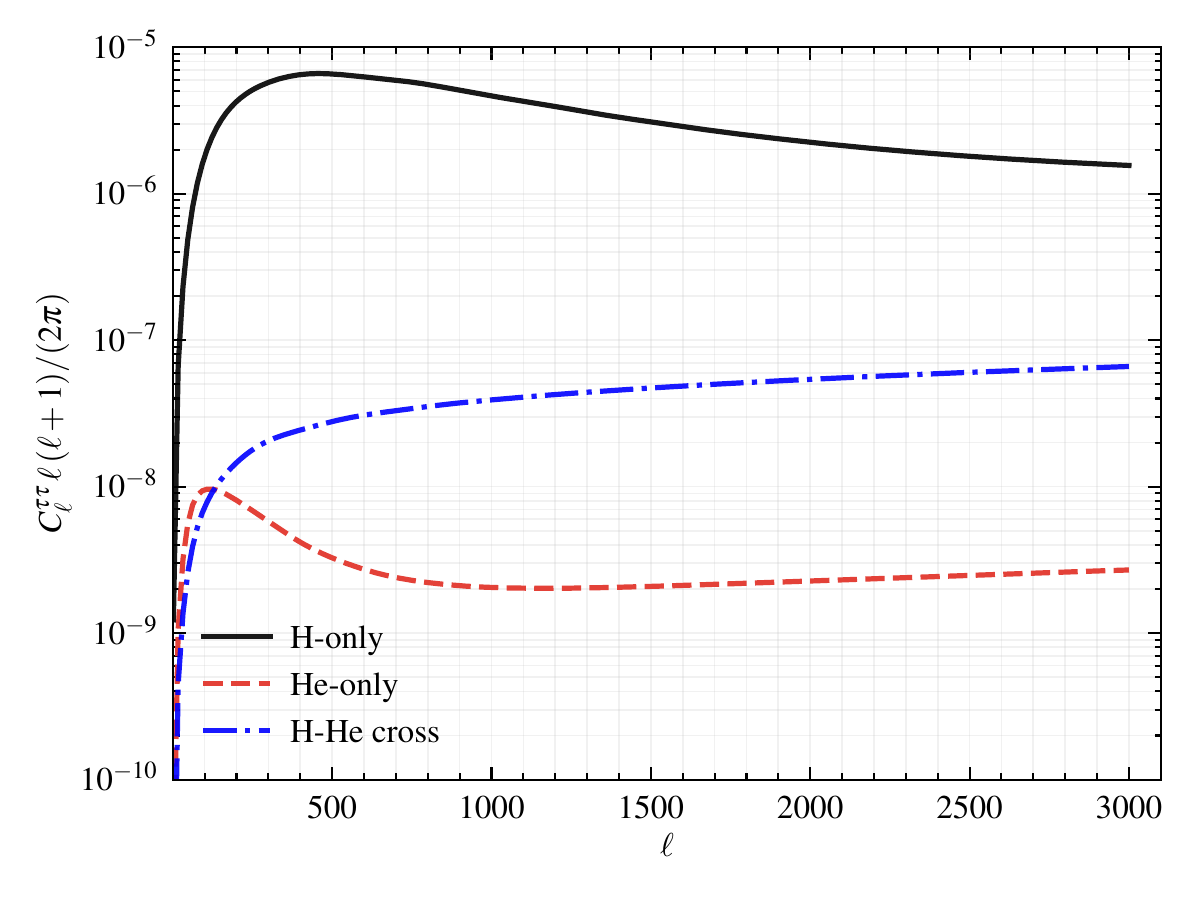} 
    \caption{Comparison of the angular power spectrum of optical-depth fluctuations $C_\ell^{\tau\tau}$ as described in Eq.~\eqref{eq:cl_tautauall}, where H-only (solid black), He-only (dashed red), and the H-He cross terms (dot-dashed blue) are given by the first, second, and third term of the equation, respectively.
}
    \label{fig:combined_tau_tau}
\end{figure}

To understand the relative weight of each of the separate terms in Eq.~\eqref{eq:cl_tautauall}, Fig.~\ref{fig:combined_tau_tau} plots each contribution as a function of multipole $\ell$ for the chosen set of fiducial parameters presented in Tab.~\ref{tab:set_of_params}.
Relative suppression of He-only and H-He cross terms is primarily sourced by the factor $f_{\rm He}$, which manifests a significantly lower abundance of helium compared to hydrogen.
While the He-only term is slightly higher than the H-He cross term at lower $\ell$, the latter dominates by about an order of magnitude at smaller scales due to comparably greater hydrogen fraction.

\begin{figure}[h!]
    \centering
    \includegraphics[width=1.0\textwidth]{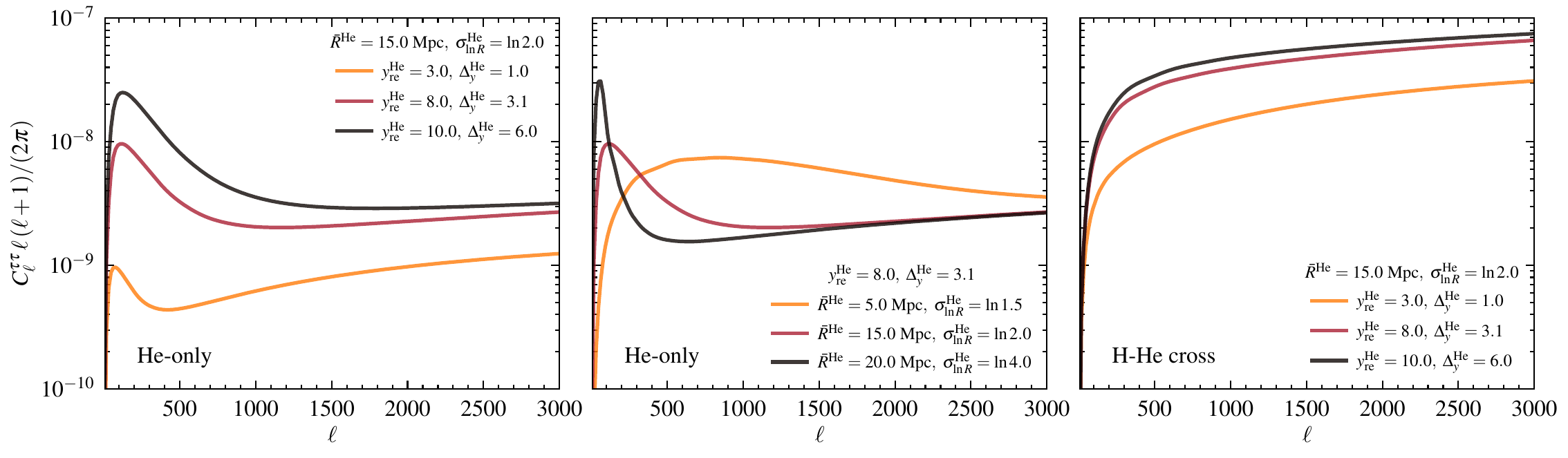} 
    \caption{Angular power spectrum of the optical-depth fluctuations $C_\ell^{\tau\tau}$ for various model parameters characterizing patchy helium (He) reionization for the He-only (H-He cross) terms of Eq.~\eqref{eq:cl_tautauall} in the left and middle (right) panels.
    In the left and right panels, the bubble radius distribution parameters are fixed to the fiducial values while the ionization fraction function parameters are varied.
    In the middle panel, the ionization fraction function parameters are fixed to the fiducial values while the bubble radius distribution parameters are varied.
    In each case, the red line shows the results for the set of fiducial parameters.
}
    \label{fig:varied_He_and_cross}
\end{figure}

To elucidate the influence of specific fiducial helium reionization parameters on the helium-related terms in Eq.~\eqref{eq:cl_tautauall}, Fig.~\ref{fig:varied_He_and_cross} presents the relevant components of $C_{\ell}^{\tau\tau}$ under different reionization scenarios. The figure's three panels highlight the maximum deviations in the $\tau$ auto-correlation within the parameter ranges specified in Tab.~\ref{tab:set_of_params}.

The left-most panel displays the minimum (orange) and maximum (black) possible `He-only' contributions considered, holding the bubble parameters $\bar{R}^{\rm He}$ and $\sigma_{\ln R}^{\rm He}$ constant while varying the ionization fraction parameters. 
Similarly, the middle panel shows the impact of the assumed He bubble size distribution by displaying the He-only contributions arising from a $P(R)$ strongly preferring smaller bubbles (orange) and one that more loosely prefers larger bubbles (black). 
This is achieved by holding the ionization parameters $y_{\rm re}^{\rm He}$ and $\Delta_y^{\rm He}$ constant while varying the bubble parameters to their extremes. Finally, the third panel displays the possible contributions from the `H-He Cross' term arising from the widest variations in $y_{\rm re}^{\rm He}$ and $\Delta_y^{\rm He}$ allowed by Tab.~\ref{tab:set_of_params}, similar to the left-most plot. In all three cases, the solid line in red displays the expectation from the assumed fiducial parameters, and the parameter variations considered correspond to the differing ionization fraction functions and bubble distributions plotted in Fig.~\ref{fig:He_variation}. 

The varied versions of signals displayed in Fig.~\ref{fig:varied_He_and_cross} can help understand the effects of the reionization model parameters on the optical-depth power spectrum. The left- and right-most panels indicate that shorter epochs of reionization lead to lower amplitude signals. The variation in the width of the low-$\ell$ peak in the left-most panel can also be attributed to the duration of reionization - the 1-bubble `shot-noise' term dominates for a larger redshift range leading to increased signal across a wider range of $\ell$ values. The middle panel displays the fact that larger bubbles (on average) correspond to peaks at smaller $\ell$, since the effective scale of patchy-correlations is increased. The dependence of the signal on $\sigma_{\ln R}^{\rm He}$ is more complicated. The effective scale of `patchy' correlations has a non-trivial dependence on $\sigma_{\ln R}^{\rm He}$, previously explored in Ref.~\cite{Mortonson:2006re}.
Ultimately, these extremal signals $C_{\ell}^{\tau\tau}$ encompass the variations considered in our forecasts in Sec.~\ref{sec:forecasts}.

\subsection{Angular cross-power between the galaxy and ionized electron fluctuations}\label{sec:Cl_taug}

Looking at Eq.~\eqref{eq:cl_tautauall} and Fig.~\ref{fig:combined_tau_tau}, one might naturally expect the signal from helium reionization to be washed-out by the (relatively) larger hydrogen counterpart in the $\tau$ auto-correlation measurement. However, noting that the ionized helium bubbles are expected to form around AGN/quasars, which trace the distribution of galaxies on large scales, we can amplify the helium signal by cross-correlating optical depth measurements with galaxy survey data. This cross-correlation takes the following functional form:
\begin{equation}\label{eq:powespec_cross_tau_gal}
C_\ell^{\tau g}=\underbrace{\int_0^{\chi_*}\dd\chi\frac{\sigma_Tn_{p,0}}{a^2\chi^2}P^{\rm H}_{\Delta_e g}(\chi,k)}_{\mathrm{H-only}}
+\underbrace{\int_0^{\chi_*}\dd\chi\frac{\sigma_T f_{\rm He}n_{p,0}}{a^2\chi^2}P^{\rm He}_{\Delta_e g}(\chi,k)}_{\mathrm{He-only}}\,.
\end{equation}
Here,
\be\label{eq:powespec_cross}
P_{\Delta_e g}^\theta(\chi,k)=[\bar{x}^\theta_e-(1-\bar{x}^\theta_e)\ln(1-\bar{x}^\theta_e)b^{\theta}\,I^\theta(k)]b_g(z) P(\chi,k)\,,
\ee
where $\bar{x}_e^\theta(z)$ similarly varying from zero to unity, $b_g(z)=b_{g,0}(1+z)$ is the galaxy bias, and we use the Limber approximation as in Eq.~\eqref{eq:cl_tautauall}. We show the cross-correlation signal in the absence of this approximation in Appendix~\ref{sec:binned_optical_depth}. 
Note that, unlike $C_{\ell}^{\tau\tau}$, our model for $C_\ell^{\tau g}$ consists only of a 2-bubble-like term.
The resulting effect is that $C_\ell^{\tau g}$ is a (relatively) large-scale probe.
While it can appreciably boost the signal, it may not be able to carry sufficient information regarding the small-scale features of He reionization, as we will discuss in Sec.~\ref{sec:forecasts}.
Based on assumptions about the small-scale distribution of galaxies, one can model a non-zero 1-bubble-like term for the $C_\ell^{\tau g}$ signal. However, given the uncertainties on models of the small-scale galaxy power spectrum, a detailed exploration of this effect is left to future work.

\begin{figure}[h]
    \centering
    \includegraphics[width=0.65\textwidth]{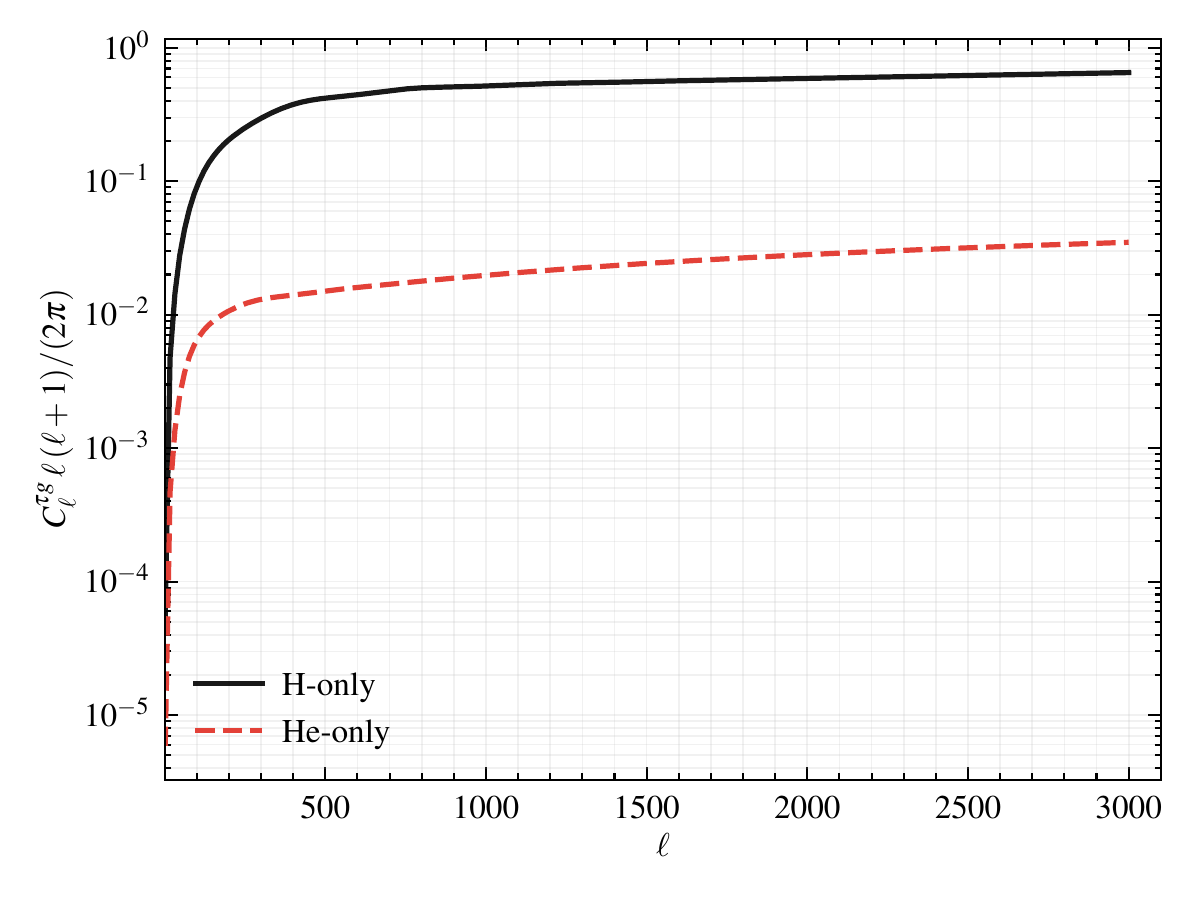} 
    \caption{Comparison of the angular cross-power spectrum between the galaxy and the optical-depth field $C_\ell^{\tau g}$ based on our fiducial hydrogen (H) and helium (He) reionization models, as described in Eq.~\ref{eq:powespec_cross_tau_gal} and~\ref{eq:powespec_cross}. The black, solid curve corresponds to the contribution from the `H-only' term to the total $C_{\ell}^{\tau g}$ signal [see  Eq.~\eqref{eq:powespec_cross_tau_gal}]. Similarly, the red, dashed curve corresponds to the contribution from the `He-only' term.
}
    \label{fig:combined_tau_gal}
\end{figure}

Figure~\ref{fig:combined_tau_gal} depicts the relative contribution from each of the terms in Eq.~\eqref{eq:powespec_cross_tau_gal}. Although it appears as if, once again, the H term is far more dominant than the He contribution, it is important to note that at the redshifts relevant to galaxy observations ($z \lesssim 5$) the power in the `H-only' term is coming from the correlation between the matter and galaxy distribution on large scales. That is, this signal varies smoothly at late-times and does not contain information on the `patchy' morphology of hydrogen reionization. On the other hand, although the helium signal is subdominant in absolute terms, it is actually this signal that causes significant redshift evolution of the $\tau-g$ signal at late times.
Furthermore, both the plotted curves represent the \textit{redshift-integrated} contribution of each term to the total cross-correlation signal. In contrast, the SNR and information-matrix analyses presented in subsequent sections leverage the \textit{redshift-binned} version of the $\tau-g$ signal. In other words, our analysis accounts for the fact that the relative contribution of the `He-only' term is much more prominent depending on the redshift-bin and the scales in consideration.
Therefore, the binned SNR and information-matrix analysis performed in this work, which accounts for the tomographic measurement of galaxy distributions, shows that most of the constraining power in the measurement of parameters characterizing helium reionization is sourced by this cross-correlation.

\begin{figure}[t]
    \centering
    \includegraphics[width=1.0\textwidth]{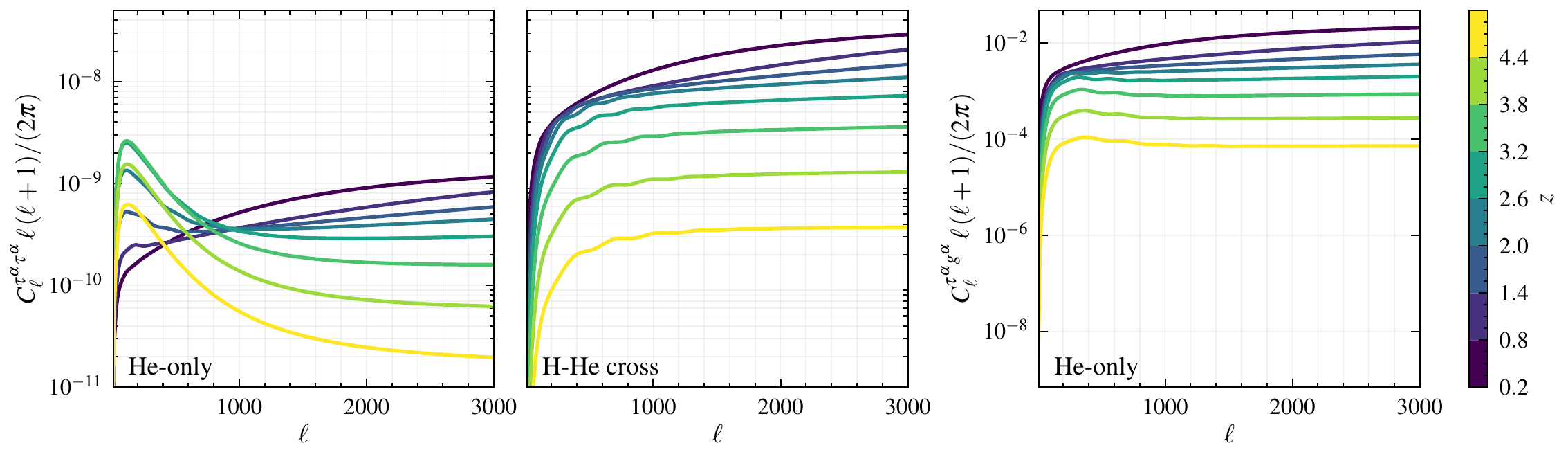} 
    \caption{Comparison of the \textit{binned} angular power spectrum of the optical depth for He-only and H-He cross components (left and middle panels, respectively) and the angular cross-power between the galaxy and the optical-depth fluctuations for the He-only component (right panel).
    Different colors correspond to the different redshift bins.
    The limits of the colors in the color map are consistent with the limits of the redshift bins used.
}
    \label{fig:binned_combined}
\end{figure}

Finally, Fig.~\ref{fig:binned_combined} displays the contributions of the He and H-He cross terms from different redshift bins to the line-of-sight integrated optical depth signal $C_{\ell}^{\tau\tau}$ and the cross-correlated signal $C_{\ell}^{\tau g}$. 
These contributions are computed for 8 equal-width redshift bins between $z = 0.2$ and $z = 5.0$. 
The contribution from each independent bin $[z_{\rm low}, z_{\rm high}]$ calculated by performing the line-of-sight integrations [Eqs.~\eqref{eq:cl_tautauall}~and~\eqref{eq:powespec_cross_tau_gal}] from lower bound $\chi(z_{\rm low})$ to upper bound $\chi(z_{\rm high})$.\footnote{A detailed description of the calculation of the redshift binned power spectra, $C_{\ell, \alpha \alpha}^{\tau \tau}$ and $C_{\ell, \alpha \alpha}^{\tau g}$, can be found in Appendix~\ref{sec:binned_optical_depth}.}

The left-most plot of Fig.~\ref{fig:binned_combined} shows the binned contributions to the `He-only' term in the model for $C_{\ell}^{\tau\tau}$. 
As expected, the signal is dominated by redshift bins spanning $5 \gtrsim z \gtrsim 1.5$, which correspond to the `patchy epoch' of 
helium reionization as set by the fiducial parameters (cf.~Fig.~\ref{fig:ion_fraction}).
Note also that while the patchy epoch dominates the signal at relatively large scales ($\ell \lesssim 800$), the angular power spectrum is dominated by lower redshift bins at smaller scales ($\ell \gtrsim 800$).
The middle panel displays the contributions from the same set of bins to the `H-He cross' term. 
In contrast to the `He-only' contributions, this plot indicates that bins at later redshifts contribute more to the cross signal independent of the scale. 
This behaviour can be attributed to the fact that our fiducial model for reionization predicts that the patchy-hydrogen and -helium power spectra approximate to the nonlinear matter power spectrum $P(\chi, k)$ at late times. 
In the cross term, at the displayed redshifts, $\bar{x}_e^{\rm H} \approx 1$ and therefore we integrate $P(\chi, k)$ modulated by $\bar{x}_e^{\rm He}$ over increasingly larger redshift bins (in $\chi$-space) at later times, explaining the increasing power with decreasing redshift. 
Similar behaviour can be seen in the right-most plot as well, which depicts the binned contributions to the `He-only' term in the $C_{\ell}^{\tau g}$ signal. 
This is once again explained by $\bar{x}_{e}^{\rm He} \rightarrow 1$, modulating the galaxy power spectrum $b_gP(\chi, k)$, integrated over increasingly large redshift bins at later times. The three plots clearly depict the power of the cross-correlation in absolute terms as well, with the helium contribution to $C_{\ell}^{\tau g}$ being $\sim 4$ orders of magnitude larger than both the helium-related terms in $C_{\ell}^{\tau \tau}$. Therefore, the cross correlation of the optical-depth field with the large-scale galaxy distribution will be an integral tool in the characterization of helium reionization.

Calculating the power spectra described above can be computationally exhaustive, especially if a wide range of parameter values and experiment configurations are considered.
To improve the efficiency of the computations, we utilize the Python package \texttt{numba}~\citep{10.1145/2833157.2833162, lam2015numba}, which is an open-source just-in-time compiler that translates a subset of Python and \texttt{NumPy} into fast machine code.
This shortens the time required to evaluate the power spectra by up to two orders of magnitude, allowing us to produce high-resolution binned spectra of the modeled signal $C_{\ell}^{\tau\tau}$ and $C_{\ell}^{\tau g}$ for varying fiducial parameter scenarios.

\subsection{Effects on the CMB}\label{sec:effects_on_CMB}

Having modelled the integrated optical-depth signal in terms of reionization parameters, we turn our attention to calculating the observable effects of patchy reionization on the CMB temperature and polarization fluctuations. Most generally, the
patchy reionization leads to three effects on the CMB:
\begin{itemize}
    \item Screening of the CMB photons, where photon intensity gets multiplied by a factor $e^{-\tau(\nhat)}$.
    \item Generation of new polarization due to Thomson scattering of CMB photons off electrons in ionized bubbles. 
    \item The kinetic Sunyaev Zel'dovich (kSZ) effect generated from the radial motion of reionization bubbles relative to the observer. 
\end{itemize}
In this work, we focus on the first two effects. The kSZ effect, which requires careful modeling of small-scale electron fluctuations, also contains valuable cosmological information~\citep[see, e.g.,][]{Smith:2016lnt,Ferraro:2018izc}, which we study in an upcoming work.

The next step, therefore, is to propagate the effect of Thomson scattering from reionization on to the CMB polarization and temperature maps. We start by expressing the observed polarization signal along the line-of-sight as follows:
\be\label{eq:losP}
(Q\pm iU)(\nhat)=\int_0^{\infty}\dd\chi \frac{\mathrm{d}\tau}{\mathrm{d}\chi} e^{-\tau(\chi\nhat)}S^{\pm}_{\rm pol}(\chi\nhat)\,,
\ee
where $Q$ and $U$ are the standard Stokes parameters, $\chi \equiv \chi(z)$ is the comoving distance to redshift $z$, $S^{\pm}_{\rm pol}(\chi\nhat)=-\sqrt{6}/10\sum_m\,_{\pm 2} Y_{2m}(\nhat)a^T_{2m}(\chi\nhat)$ is the local temperature quadrupole that the electron sees, $a_{2m}^T(\chi\nhat)$ are the temperature quadrupole moments, and $\,_{\pm 2} Y_{2m}(\nhat)$ are the spin $\pm2$ spherical harmonics. 
Similar to Ref.~\citep{Dvorkin:2008tf}, we parameterize the effect of reionization on CMB polarization anisotropies as
\be\label{eq:patchylosP}
\begin{split}
(Q\pm iU)(\nhat)=&(Q\pm iU)_0(\nhat)+\sigma_T n_{p,0}\int\frac{\dd \chi}{a^2}x_e(\chi\nhat)(Q\pm i U)_1(\chi\nhat)\,,
\end{split}
\ee
where $(Q\pm iU)_0(\nhat)$ is the polarization signal in the absence of these effects, $x_e(\chi\nhat)$ is the spatial fluctuations in the total ionized electron fraction introduced above, and
\be
(Q\pm i U)_1(\chi\nhat)=e^{\tau(\chi)}S^{\pm}_{\rm pol}(\chi\nhat)-\int_\chi^\infty\dd\chi'\frac{\dd\tau}{\dd\chi'}e^{-\tau(\chi)}S_{\rm pol}^{\pm}(\chi\nhat)\,,
\ee
is the modulation to the polarization sourced by patchy reionization. 
Here, $\tau(\chi)$ is the spatially-averaged mean optical depth at comoving distance $\chi$. Throughout, we represent the CMB signals in redshift-binned form, introducing $N$ redshift bins covering the patchy epoch of reionization. The polarization signal takes the form
\be
(Q\pm i U)(\nhat)=(Q\pm iU)_0(\nhat)+\sum\limits^N_\alpha \Delta\tau^\alpha(\nhat)(Q\pm iU)_1(\chi_\alpha\nhat)\,,
\ee
where $\Delta\tau^\alpha$ is the redshift-bin--averaged optical-depth fluctuation. The effect of reionization on the CMB temperature anisotropy can also be written in a form similar to polarization, satisfying 
\be
T(\nhat)=T_0(\nhat)+\sum\limits^N_\alpha \Delta\tau^\alpha(\nhat) T_1(\chi_\alpha\nhat)\,,
\ee
with
\be
T(\nhat)=\int_0^\infty \dd\chi S_T(\chi\nhat)\,,
\ee
where $S_T(\chi\nhat,\tau)$ is a function of local quantities, dependent on the optical depth, albeit in a more complicated way due to multiple contributing effects. The modulation $T^\alpha_1(\chi\nhat)$ in this case can be expressed with a functional integral of the form
\be
T_1^\alpha(\chi\nhat)=\int_{\chi_\alpha}^\infty \frac{\delta T(\chi\nhat)}{\delta\tau(\chi)}\,,
\ee
and can be found in Ref.~\citep{Dvorkin:2008tf}, for example. 

Since $Q$ and $U$ are not coordinate invariant, it is more convenient to re-cast these maps in terms of the scalar $E$ and $B$ fields using spin-raising and -lowering operators. The harmonic coefficients of the CMB $E$- and $B$-mode maps satisfy 
\be
\begin{split}
a^{E}_{\ell m }&\equiv\frac{1}{2}(\,_{+2}a_{\ell m}+\,_{-2} a_{\ell m})\,,\\
a^{B}_{\ell m }&\equiv\frac{1}{2i}(\,_{+2}a_{\ell m}-\,_{-2} a_{\ell m})\,,
\end{split}
\ee 
where 
\be
\,_{\pm2} a_{\ell m}= \int\dd^2\nhat(Q\pm iU)(\nhat)\,_{\pm2}Y_{\ell m}(\nhat).
\ee
The harmonic coefficients for the $E$-mode polarization, $a_{\ell m}^{E_0}$ and $a_{\ell m}^{E_1^\alpha}$, then satisfy 
\be 
\begin{split}
a_{\ell m}^{E_0}&=\frac{1}{2}(\,_{+2}a^{E_0}_{\ell m}+\,_{-2} a^{E_0}_{\ell m})\,,\\
a_{\ell m}^{E_1^{\alpha}}&=\frac{1}{2}(\,_{+2}a^{E_1^{\alpha}}_{\ell m}+\,_{-2} a^{E_1^{\alpha}}_{\ell m})\,,
\end{split}
\ee 
where $a^{E_0}_{\ell m}=\int\dd^2\nhat(Q\pm i U)_{0}(\nhat)\,_{\pm2}Y_{\ell m}(\nhat)$ for the homogeneous contribution and $a^{E_1^\alpha}_{\ell m}=\int\dd^2\nhat(Q\pm i U)_{1}(\chi_\alpha\nhat)\,_{\pm2}Y_{\ell m}(\nhat)$ for the contribution from patchy reionization. Similarly, the temperature harmonic coefficients from patchy reionization satisfy
\be 
T_1(\chi_\alpha\nhat)=\sum\limits_{\ell m} a_{\ell m}^{T_1(\chi_\alpha)}Y_{\ell m}(\nhat)\,,
\ee
and \be T_0(\nhat)=\sum\limits_{\ell m}a_{\ell m}^{T_0}Y_{\ell m}(\nhat)\,.\ee 
In order to calculate the effect of patchy reionization on the observed CMB spectra, which we use to define optical-depth estimators in the next section, it is sufficient to calculate the angular cross-correlation functions from these coefficients, which are given by   
\be
C_\ell^{E_0E_1^\alpha}\delta_{\ell \ell'}\delta_{mm'}=\langle a^{E_0}_{\ell m}a^{E_1^\alpha}_{\ell' m'}\rangle\,,
\ee
\be
C_\ell^{T_0E_1^\alpha}\delta_{\ell \ell'}\delta_{mm'}=\langle a^{T_0}_{\ell m}a^{E_1^\alpha}_{\ell' m'}\rangle\,,
\ee
\be
C_\ell^{T_0T_1^\alpha}\delta_{\ell \ell'}\delta_{mm'}=\langle a^{T_0}_{\ell m}a^{T_1^\alpha}_{\ell' m'}\rangle\,,
\ee
where $\delta_{ij}$ is the Dirac delta function.
The specific form of the coefficients $E_1^\alpha$, $T_1^\alpha$ are given in Ref.~\citep{Dvorkin:2008tf}.

\section{Patchy optical depth estimator}\label{sec:tau_estimator}

The three effects described in Sec.~\ref{sec:effects_on_CMB} introduce a statistical anisotropy in the small-scale CMB temperature and polarization, which can be used to reconstruct the underlying fluctuations of the optical depth $\Delta \tau$. At leading order in $\Delta \tau$, the cross-correlation of two CMB fields can be written as 
\be\label{eq:cross_CMB}
\begin{split}
\langle a^X_{\ell m}a^Y_{\ell' m'}\rangle=&(-1)^{m}C_{\ell}^{X_0Y_0}\delta_{\ell\ell'}\delta_{mm'}+\sum\limits^N_\alpha\!\sum\limits_{LML'M'}\!\!\Delta\tau^\alpha_{L'M'}\langle a^{X_0}_{LM} a^{Y_1^\alpha}_{L'M'}\rangle\int\dd^2\nhat Y_{\ell m}^*(\nhat)Y_{LM}(\nhat)Y_{L'M'}(\nhat)\\
=&(-1)^{m}C_{\ell}^{X_0Y_0} \delta_{\ell\ell'}\delta_{mm'} +\sum\limits^N_\alpha\sum\limits_{ML}\Delta\tau^\alpha_{LM}\Gamma_{\ell\ell'L}^{X_0Y^\alpha_1}
\begin{pmatrix}
\ell & \ell' & L\\
m & m' & M
\end{pmatrix}\,,
\end{split}
\ee
where $\Delta \tau^\alpha$ is the redshift-bin averaged optical-depth fluctuations, for which a \textit{biased} minimum variance estimator can be written as 
\be
\begin{split}
&\widehat{\Delta\tau_{\ell m}^{\alpha}}\,{\rm(biased)}=A_{\ell,\alpha\alpha}^{XY}(-1)^m\!\!\!\!\sum\limits_{LL'MM'}\!\!\begin{pmatrix}
L & L' & \ell \\
 \!M & \!M' & m
\end{pmatrix}\Gamma^{X_0Y^\alpha_1}_{LL'\ell}\frac{a^X_{LM}a^Y_{L'M'}}{C_{L}^{XX,\rm obs}C_{L'}^{YY,\rm obs}}\,,
\end{split}
\ee
where
\be
A^{XY}_{\ell,\alpha\beta}=\left[\frac{1}{2\ell+1}\sum\limits_{\ell'L}\frac{\Gamma^{X_0Y^\alpha}_{LL'\ell}(\Gamma^{X_0Y^\beta_1}_{LL'\ell})^*}{C_{L}^{XX,\rm obs}C_{L'}^{YY,\rm obs}}\right]^{-1}\,.
\ee
The above estimator has a biased optical-depth reconstruction noise given by
\be
{N}_{\ell,\alpha\beta}^{\tau\tau;XY}\,(\rm biased) = \frac{A^{XY}_{\ell,\alpha\alpha} A_{\ell,\beta\beta}^{XY}}{A^{XY}_{\ell,\alpha\beta}}\,.
\ee
We can also define an \textit{unbiased} quadratic estimator for the optical depth as 
\be
\widehat{\Delta\tau_{\ell m}^{\alpha}}=(\boldsymbol{R}^{-1})_{\alpha\gamma}\widehat{\Delta\tau_{\ell m}^{\gamma}}{{\,(\rm biased)}}\,,
\ee
which satisfies $\langle\widehat{\Delta\tau_{\ell m}^{\alpha}}\rangle=\Delta\tau_{\ell m}^{\alpha}$, where $\boldsymbol{R}$ is a rotation matrix that de-biases the reconstructed optical depth, whose elements can be found to satisfy
\be\label{eq:rotation}
R_{\alpha\beta}^{XY}=\frac{A^{XY}_{\ell,\alpha\alpha}}{A^{XY}_{\ell,\alpha\beta}}\,.
\ee
The reconstruction noise for the unbiased estimator can then be calculated as
\be\label{eq:noise_unbiased}
N_{\ell,\alpha\beta}^{\tau\tau;XY}=(\boldsymbol{R}^{-1})_{\alpha\gamma}^{XY}(\boldsymbol{R}^{-1})_{\beta\delta}^{XY}\,\,\tilde{N}_{\ell,\gamma\delta}^{\tau\tau;XY}\,\,\delta_{\ell\ell'}\delta_{mm'}\,,
\ee
and we define a minimum-variance reconstruction noise by combining these estimators as
\be\label{eq:minvar}
{{N}_{\ell,\alpha\beta}^{\tau\tau;\rm MV}}=(\sum_{XY}({\boldsymbol{N}_{\ell}^{\tau\tau;XY}})^{-1})^{-1}_{\alpha\beta}\,,
\ee 
where $X,Y\in\{T,E,B\}$ and $\alpha,\beta$ span the redshift-bin indices. {Note that we omit correlations between different redshift bins when calculating the signal from Eq.~\eqref{eq:cross_CMB}, as we find these terms to be small and inconsequential to our minimum-variance reconstruction noise. In close analogy with CMB lensing reconstruction, the covariance of the reconstruction noise spectra also includes off-diagonal cross-correlation terms between different estimators. We surmise that taking into account these off-diagonal covariance terms may increase the minimum-variance reconstruction noise up to around a factor of $\mathcal{O}(0.1)$, similar to the case of lensing reconstruction.
We leave studying the impact of off-diagonal contributions to the optical-depth reconstruction noise covariance to future work~\citep[see, e.g.,][for calculations corresponding to lensing quadratic estimator]{Hotinli:2021umk,Hotinli:2020adc}.} We show our calculations of different optical-depth estimators in Appendix~\ref{sec:estimator_detail}.

\section{Forecasts}\label{sec:forecasts}

Building on the target optical depth signal outlined in Sec.~\ref{sec:He_reionization_model} and the measurement methodology and noise considerations for reconstructing this field presented in Sec.~\ref{sec:tau_estimator}, we now advance to forecasting measurement accuracy for forthcoming CMB and LSS surveys. These forecasts are based on the standard \textit{Planck} 2018~\citep{Planck:2018vyg} $\Lambda\textrm{CDM}$ cosmology, employing the six parameters specified in Table~\ref{table:cosmo_fiducial}. 
Sec.~\ref{sec:experiments} provides an overview of experimental setups, including beam characteristics and noise assessments. In Sec.~\ref{sec:SNR_forecasts}, we analyze the auto-correlation SNR of the reconstructed patchy optical depth and the optical-depth-galaxy cross-correlation SNR (Eqs.~\eqref{eq:SNRauto} and~\eqref{eq:SNRcross}), presenting results for various experimental scenarios and exploring the impact of He reionization parameter variations. Finally, Sec.~\ref{sec:fisher_forecasts} delves into the measurability of H and He reionization parameters, assessing potential parameter degeneracies and the influence of varying He reionization parameters.

\subsection{Experiments}\label{sec:experiments}

\begin{table}[t!]
\centering
\caption{Fiducial cosmological parameters for the 6-parameter $\Lambda$CDM model considered in our calculations throughout this paper matching \textit{Planck} 2018~\cite{Planck:2018vyg}.} 
\label{table:cosmo_fiducial}
\vspace{10pt}
\begin{minipage}{.65\textwidth} 
\renewcommand{\arraystretch}{1.3} 
\centering
\begin{tabular}{ll}
\hline \hline
   Parameter & Fiducial Value \\
\hline
   Cold dark matter density ($\Omega_c h^2$) & 0.120 \\
   Baryon density ($\Omega_b h^2$) & 0.022 \\
   Angle subtended by acoustic scale ($\theta_s$) & 0.010410 \\
   Optical depth to recombination $(\tau)$ & 0.060 \\
   Primordial scalar fluctuation amplitude ($A_s$) & $2.196\!\times\!10^{-9}$ \\
   Primordial scalar fluctuation slope ($n_s$) & 0.965 \\ \hline \hline
\end{tabular}
\end{minipage}
\end{table}

\begin{table}[h!]
\caption{{\it Inputs to ILC noise for the baseline CMB configurations:} 
The beam and temperature noise RMS parameters are chosen to roughly match CMB-S4~\citep{CMB-S4:2016ple} and CMB-HD~\citep{CMB-HD:2022bsz}. 
We model the CMB noise as shown in Eq.~\eqref{eq:instrument_noise1}. In both cases, we account for the degradation due to Earth's atmosphere by including the CMB red-noise with $\ell_{\rm knee}=100$ and $\alpha_{\rm knee}=3$. 
The polarization noise satisfies $\Delta_E=\Delta_B=\sqrt{2}\Delta_T$.}
\label{tab:beamnoise}
\begin{center}
\begin{minipage}{\textwidth}
\renewcommand{\arraystretch}{1.3} 
\centering
\begin{tabular}{lcccc}
\hline \hline
\  \ & \multicolumn{2}{c}{Beam FWHM} \ \ & \multicolumn{2}{c}{Noise RMS} \\
& \multicolumn{2}{c}{} & \multicolumn{2}{c}{($\mu$K-arcmin)} \\ 
\hline
& S4 & HD & S4 & HD \\ \hline
39 GHz  \   \             & $5.1'$                  & $36.3''$       \    \                          & 12.4                     & 3.4                     \\
93 GHz \   \              & $2.2'$                  & $15.3''$        \     \                         & 2.0                     & 0.6                     \\
145 GHz   \   \           & $1.4'$                  & $10.0''$       \     \                         & 2.0                     & 0.6                     \\
225 GHz \   \            & $1.0'$                  & $6.6''$       \      \                          & 6.9                     & 1.9                     \\
280 GHz  \  \             & $0.9'$                  & $5.4''$      \     \                           & 16.7                    & 4.6                     \\ \hline \hline
\end{tabular}
\end{minipage}
\end{center}
\end{table}

We model atmospheric and instrumental noise contributions to the CMB temperature and polarization at a given frequency as 
\be\label{eq:instrument_noise1}
N_\ell^{\rm TT}=\Delta_T^2\exp\left(\frac{\ell(\ell+1)\theta^2_{\rm FWHM}}{8\ln2}\right)\left[1+(\ell_{\rm knee}/\ell)^{\alpha_{\rm knee}}\right]\,,
\ee
assuming $N_\ell^{EE}=N_\ell^{BB}=2N_\ell^{\rm TT}$, where $\Delta_T$ is the detector RMS noise, and $\theta_{\rm FWHM}$ is the Gaussian beam full width at half maximum.
The term inside the square-brackets in Eq.~\eqref{eq:instrument_noise1} corresponds to the `red' noise due to Earth's atmosphere, parameterised by the parameters $\ell_{\rm knee} = 100$ and $\alpha_{\rm knee} = 3$. 
We define our choices for these parameters in Table~\ref{tab:beamnoise}, which we set to match the ongoing and upcoming CMB surveys---CMB-S4~\citep{Abazajian:2016yjj,Abazajian:2019eic,1907.04473} and CMB-HD~\citep{Sehgal:2019ewc,CMB-HD:2022bsz}.
Note also that in our analysis we take $\ell_{\rm max}$, the smallest scales probed by CMB, equal to $10^4$ both for CMB-S4 and CMB-HD. A CMB-HD--like survey in principle can access smaller scales which could improve our results.

To account for residual contamination from foregrounds to the CMB, we consider Poisson and clustered cosmic infrared background (CIB), as well as the thermal Sunyaev-Zel'dovich (tSZ)  foreground, which we calculate following Refs.~\citep{Madhavacheril:2017onh, Park:2013mv}. 
We omit the cross-correlation between tSZ and CIB. We include radio point sources following Ref.~\citep{Lagache19}. 
In addition to frequency-dependent contributions, we also consider black-body, late-time, and reionization kSZ and calculate the lensed CMB black-body using \texttt{CAMB}~\citep{CAMB}.
For the forecasts that follow, we assume that the galactic foreground is removed from the CMB temperature and polarization maps. To remain consistent with these assumptions, we provide forecasts assuming a sky coverage of $f_{\rm sky}=0.4$.
In what follows, we perform forecasts for both ILC-cleaned CMB and `black-body' CMB, where for the latter, we omit contributions from frequency-dependent foregrounds but include residual white noise after ILC cleaning.

For the galaxy density, we consider a galaxy catalogue with specifications anticipated to match the photometric LSST survey~\citep{2009arXiv0912.0201L}. 
We approximate the galaxy density of the ``gold'' sample with $n_\text{gal}(z) = n_0[({z}/{z_0}]^2\exp(-z/z_0)/{2z_0}$, where $n_0=40~\text{arcmin}^{-2}$, $z_0=0.3$, and we take the galaxy bias as $b_g(z)=0.95(1+z)$.
While accounting for the standard anticipated photometric redshift (photo-$z$) error $\sigma_z = 0.03(1 + z)$, it is noteworthy that the galaxy redshift bin sizes chosen in this analysis—eight equal-width redshift bins within the range $z \in [0.2, 5.0]$—are considerably larger than the photometric redshift errors at all redshifts.
As a result, we anticipate that photo-$z$ errors will not significantly impact our results. While this study does not include it, a future spectroscopic survey such as MegaMapper could, in principle, enable much finer redshift binning. This would enhance the statistical significance of the joint analysis with the CMB data.
We leave forecasting for MegaMapper to future work.

\subsection{Detection SNR of the patchy optical depth}\label{sec:SNR_forecasts}

To evaluate the statistical power of upcoming CMB and galaxy surveys in detecting optical-depth fluctuations caused by the ionized second electron of helium, we initially investigate the SNR of the reconstructed patchy optical depth.
We define the auto-correlation SNR as
\be\label{eq:SNRauto}
 {\rm SNR}^2~({\rm auto})=\sum_{\ell\ell'}\sum_{\alpha\beta\gamma\delta}C_\ell^{{{\tau}}_\alpha {{\tau}}_\beta}\textbf{cov}^{-1} \left(\tilde{C}^{{\tau}_\alpha {\tau}_\beta}_{\ell},\tilde{C}^{{\tau}_\gamma {\tau}_\delta}_{\ell'}\right) C_{\ell'}^{{{\tau}}_\gamma {{\tau}}_\delta}\,,
\ee
where the index $\alpha$ labels a redshift bin, such that $C_\ell^{{\tau}_\alpha {\tau}_\beta}$ represents the correlation of $\tau$ across two redshift bins centered at $z_\alpha$ and $z_{\beta}$.
Note that spectra with tilde represent the \textit{observed} spectra $\tilde{C}_\ell^{{\tau}_\alpha {\tau}_\beta}\equiv C_\ell^{{\tau}_\alpha{\tau}_\beta}+ N^{\tau\tau;\rm MV}_{\ell,\alpha\beta}$, including the optical-depth reconstruction noise $N^{\tau\tau;\rm MV}_{\ell,\alpha\beta}$ defined in Eq.~\eqref{eq:minvar}, and the optical-depth signal $C_\ell^{\tau_\alpha\tau_\beta}$, which we will assume to be diagonal in redshift bins.

Using the same notational conventions, we define the reconstructed optical-depth--galaxy cross-correlation SNR as
\be\label{eq:SNRcross}
{\rm SNR}^2~(\rm cross)=\sum_{\ell\ell'}\sum_{\alpha\beta\gamma\delta}C_\ell^{{\tau}_\alpha g_\beta}\textbf{cov}^{-1} \left(\tilde{C}^{{\tau}_\alpha g_\beta}_{\ell},\tilde{C}^{{\tau}_\gamma g_\delta}_{\ell'}\right) C_{\ell'}^{{\tau}_\gamma g_\delta}  \,.
\ee
In both the SNR equations, the covariance satisfies 
\be\label{eq:covariance}
\textbf{cov} \left( \tilde{C}^{X Y}_{\ell}, \tilde{C}^{WZ}_{\ell'} \right) =  \frac{\delta_{\ell\ell'}}{2\ell+1} f_\mathrm{sky}^{-1} \left(\tilde{C}_{\ell}^{XW}\tilde{C}_{\ell}^{YZ}+\tilde{C}_{\ell}^{XZ}\tilde{C}_{\ell}^{YW}\right)\,,
\ee
where $f_{\rm sky}$ is the survey coverage sky fraction.

\begin{figure}[t!]
    \centering
    \includegraphics[width=1\textwidth]{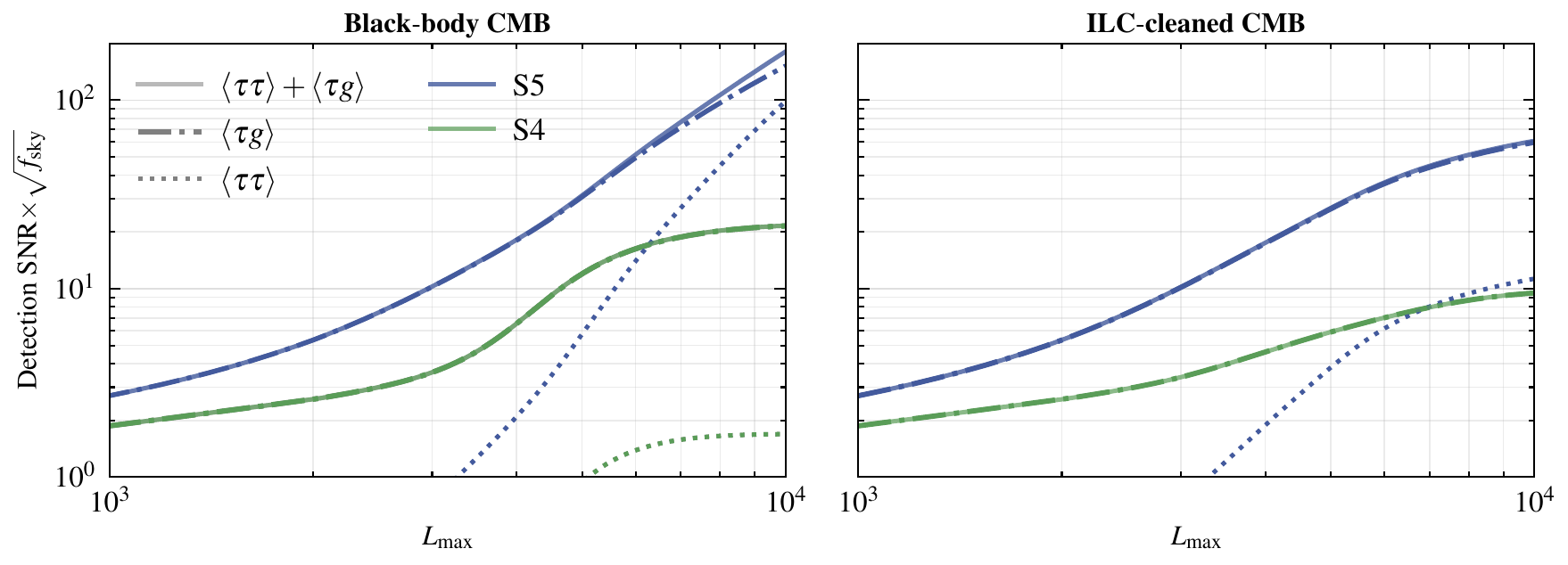} 
    \vspace*{-0.8cm}
    \caption{Detection SNR as a function of maximum multiple $L_{\rm max}$ considered, depicting the statistical power of CMB and LSS experiments (specifically, LSST) for detecting optical-depth fluctuations from the doubly-ionized helium. 
    The survey sky fraction $f_{\rm sky}$ is normalized to unity.
    The left panel corresponds to considering only the black-body contribution to the CMB. 
    The right panel corresponds to using ILC-cleaned CMB spectra, including residual frequency-dependent foregrounds as described in Sec.~\ref{sec:experiments}. 
    The dotted lines labeled $\langle \tau \tau \rangle$ correspond to the detection SNR from auto-correlation of the reconstructed $\tau$ [including both the He-only and H-He~`cross' terms in Eq.~\eqref{eq:cl_tautauall}]. 
    The dot-dashed lines labeled $\langle \tau g \rangle$ correspond to detection SNR from cross-correlating the reconstructed optical depth and observed galaxy number density with survey specifications matching LSST [He-only term in Eq.~\eqref{eq:powespec_cross_tau_gal}].
    Solid lines correspond to the combination of both signals.
    The green (labelled `S4') and blue (labelled `S5') correspond to the assumption of a CMB-S4-like and CMB-HD-like CMB experiment, respectively. 
    We find that near-future CMB and LSS experiments will have enough statistical power to detect optical-depth fluctuations from reionized helium, with detection significance reaching $\gtrsim 3 \sigma$ ($\gtrsim 10 \sigma$) from optical-depth reconstruction up to scales $L_{\rm max}\sim3000$ for CMB-S4 (CMB-HD) together with LSST, or $L_{\rm max}\sim4000$ if $f_{\rm sky}=0.4$.
    For most $L_{\rm max}$ values, the SNR is dominated by the cross-correlation of the reconstructed optical depth and galaxy field, whereas the reconstructed optical depth can be detected as a standalone probe with a CMB-HD-like survey after ILC cleaning. 
    }
    \vspace*{-0.25cm}
    \label{fig:SNR_1}
\end{figure}

\begin{figure}[h!]
    \centering
    \includegraphics[width=1\textwidth]{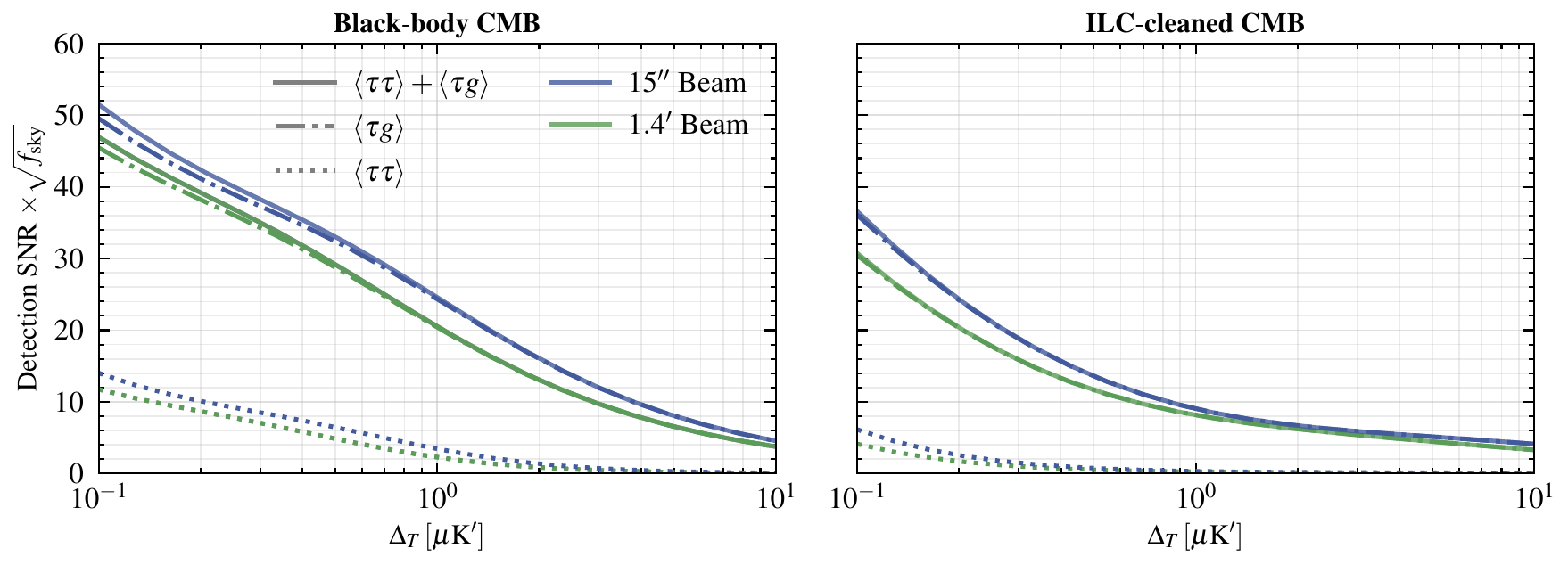} 
    \vspace*{-0.8cm}
    \caption{The dependence of the detection SNR of optical-depth fluctuations from helium reionization on the CMB white-noise RMS $\Delta_T$ in ${\rm \mu K}$-arcmin. Here, $L_{\rm max}$ is fixed to $6000$. 
    The left (right) panel corresponds to black-body (ILC-cleaned) CMB spectra. 
    The blue (green) line corresponds to a $1.4'$ ($15''$) CMB beam representing a CMB-S4-like (CMB-HD-like) telescope.
    Here, $f_{\rm sky}$ is normalized to unity.
    Detection SNR can be appropriately scaled by $\sqrt{f_{\rm sky}}$ based on the sky-coverage fraction of a given survey.
    Note that while improving the beam quality alone does not significantly enhance the SNR at a given noise level, it is important to consider that the white noise RMS $\Delta_T$ for CMB-S4 (CMB-HD) is approximately $1\mu{\rm K}$-arcmin ($0.2\mu{\rm K}$-arcmin).}
    \label{fig:SNR_2}
\end{figure}

We show the SNR forecasts for a range of experimental specifications in Figs.~\ref{fig:SNR_1} and~\ref{fig:SNR_2}.
Figure~\ref{fig:SNR_1} shows the detection SNR as a function of $L_{\rm max}$ for experiment specifications corresponding to CMB-S4 (green colored lines labelled S4) and CMB-HD (blue colored lines labelled S5). 
The dashed, dot-dashed, and solid lines correspond to SNR obtained from the the auto-correlation of the reconstructed optical-depth field $\langle \tau \tau \rangle$ [cf.~Eq.~\eqref{eq:SNRauto}], cross correlation of reconstructed optical-depth field with LSST $\langle \tau g \rangle$ [cf.~Eq.~\eqref{eq:SNRcross}], and the total of the two, respectively.
The left panel displays the resulting SNR from considering only the black-body contributions to the CMB (in addition to white noise). 
In contrast, the right panel corresponds to using ILC-cleaned CMB spectra including all foregrounds as described in Sec.~\ref{sec:experiments}. Throughout, we take $f_{\rm sky}=1$ due to the uncertainty in the sky coverage of upcoming experiments. 
The detection SNR can be straight-forwardly scaled by $\sqrt{f_{\rm sky}}$ based on the coverage of a given survey. 
For CMB-S4, the anticipated sky coverage fraction is $\sim0.4$. 
For the joint sky area surveyed by LSST and CMB-S4, $f_{\rm sky}$ is \textit{also} expected to be $\sim0.4$, while CMB-HD is conceived to cover over half of the sky~\citep{CMB-HD:2022bsz}.

We find that a joint analysis of (near future) CMB-S4 and LSST will have enough statistical power to make a high-significance ($\gtrsim3\sigma$) detection  of optical depth fluctuations sourced by the second electron in helium, once the field is reconstructed up to scales $L_{\rm max}\sim3000$ (or $L_{\rm max}\sim4000$, if $f_{\rm sky}=0.4$), following standard ILC cleaning.  
An S5 experiment like CMB-HD, moreover, could reach beyond $\gtrsim10\sigma$ detection significance for the same $L_{\rm max}$ choices. 
If the reconstructed optical depth could be utilized up to smaller scales, $L_{\rm max}\simeq10000$ for example, detection SNR for both survey combinations can be boosted by more than an order of magnitude.\footnote{Reconstructions up to high $L_{\rm max}$ will likely require modelling the 1-bubble-like term in the $\tau \times g$ cross-correlation that we have omitted in our analysis. As a result, the feasibility of this improvement will depend on the modeling uncertainties of the nonlinear contributions to galaxy distributions and the reconstructed optical depth.} 

The statistical power of these measurements depends significantly on the residual foregrounds after standard ILC cleaning, as can be seen by comparing the ILC-cleaned results (right panel of Fig.~\ref{fig:SNR_1}) to the detection SNR forecasts assuming black-body CMB (left panel of Fig.~\ref{fig:SNR_1}), which are more optimistic. We also find, for example, if frequency-dependent-foregrounds could be removed more substantially, the optical-depth fluctuations can be detected from auto-correlations with a CMB-HD-like survey at a comparable SNR to that of the $\tau$-galaxy cross-correlation.
More effective ILC-cleaning methods, such as the needlet ILC techniques~\citep[e.g.][]{Delabrouille:2008qd,McCarthy:2023hpa}, as well as more recently developing machine-learning methods~\citep{Petroff:2020fbf} may in principle improve detection prospects.
Here, we show both scenarios to demonstrate the lower and upper limits of the statistical power of our analysis, given our survey specifications. 

Figure~\ref{fig:SNR_2} further demonstrates the dependence of the detection SNR on the CMB RMS noise $\Delta_T$ in $\rm \mu K$-arcmin for a fixed choice of $L_{\rm max}$. Here, our S4 (S5) survey corresponds to a $1.4'$ ($15''$) beam.

\begin{figure}[h!]
    \centering
    \includegraphics[width=0.85\textwidth]{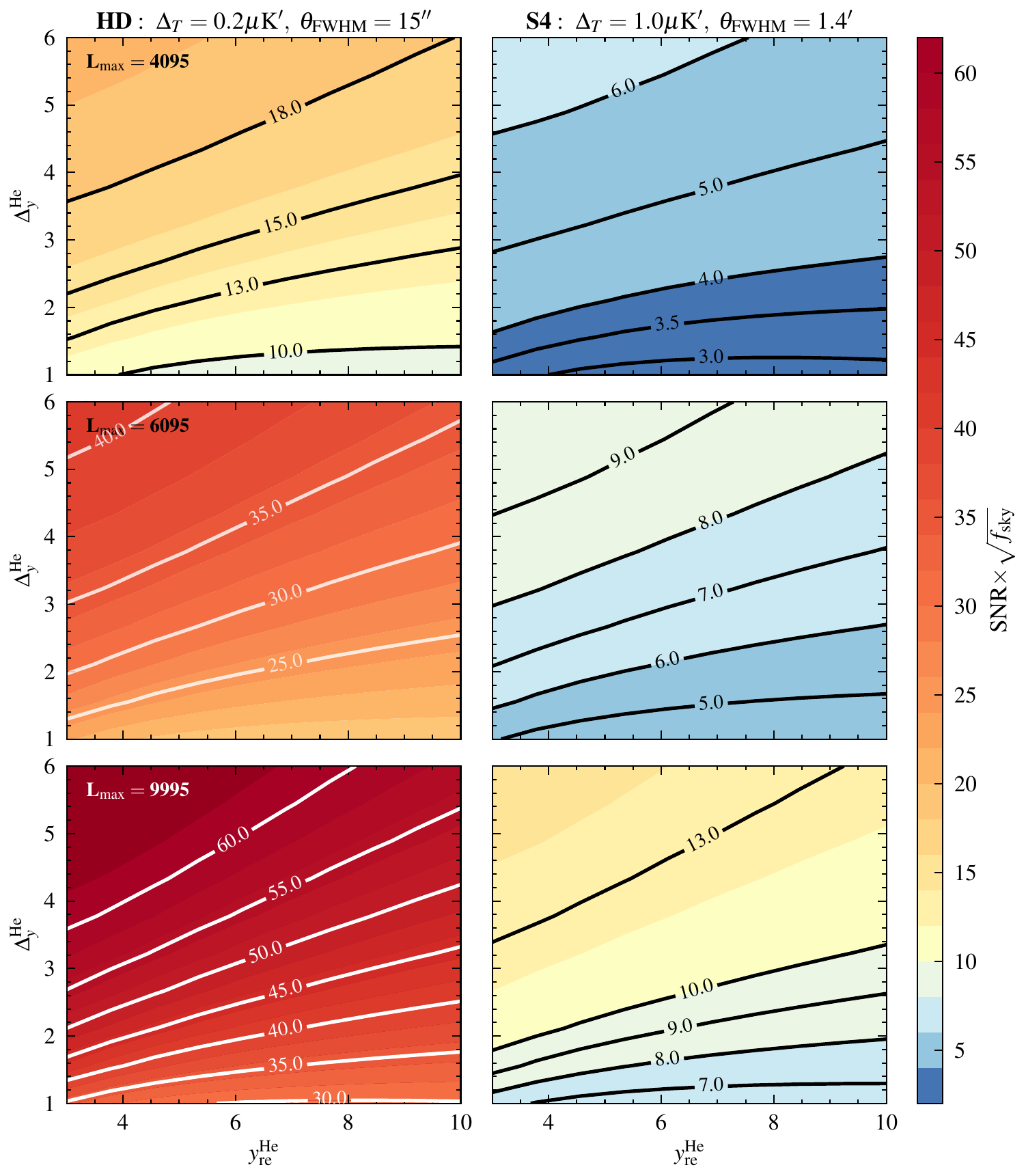} 
    \vspace*{-0.5cm}
    \caption{
    Helium reionization optical-depth detection SNR for a range of values of helium-reionization parameters $y_{\rm re}^{\rm He}$ and $\Delta_y^{\rm He}$. 
    Here, $y_{\rm re}^{\rm He}$ sets the mean-redshift of reionization while $\Delta_y^{\rm He}$ determines the duration (cf. Fig.~\ref{fig:He_variation}).
    The bubble distribution parameters are fixed to their fiducial values indicated in Tab.~\ref{tab:set_of_params}, and the sky-coverage faction $f_{\rm sky}$ is normalized to unity.
    The top, middle, and bottom rows set the maximum multipole of the reconstructed optical depth to 4095, 6095 and 9995, respectively.
    The left (right) column corresponds to a CMB-HD-like (CMB-S4-like) survey specification after ILC cleaning.
    We find that longer duration (slower transition) combined with later reionization (patchy epoch coinciding with galaxies) leads to higher SNRs.
    In principle, the SNR here can be used to put bounds on helium-reionization parameters.
d    In case the optical-depth signal from helium is not detected above $\textrm{SNR} \sim 5$ with CMB-S4 for $L_{\rm max}=4095$ (or $\sim3\sigma$ if $f_{\rm sky}=0.4$), for example, this could suggest helium reionization might have happened more rapidly, i.e.~smaller $\Delta_y^{\rm He}$, or started at a an earlier redshift, i.e.~lower $y_{\rm re}^{\rm He}$.}
    \label{fig:SNR_contour_xe}
    \vspace*{-0.5cm}
\end{figure}

\begin{figure}[h!]
    \centering
    \includegraphics[width=1\textwidth]{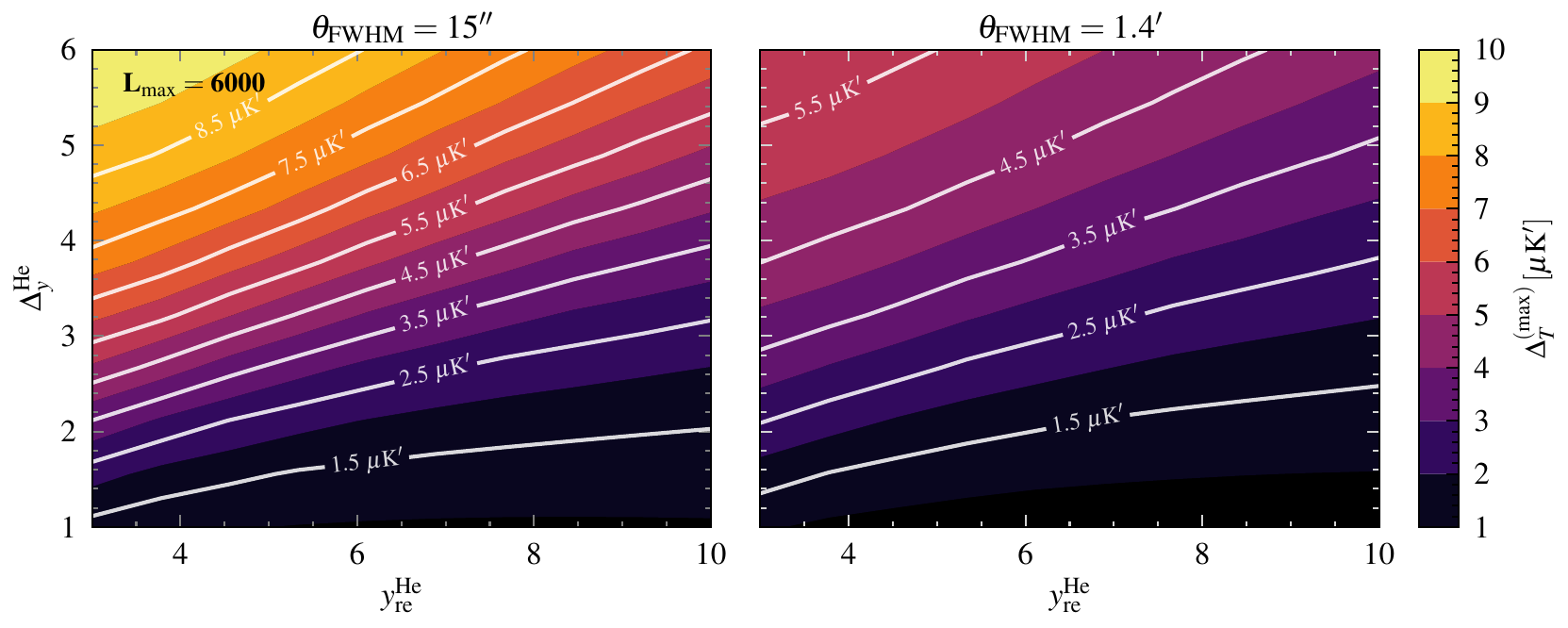} 
    \vspace*{-0.7cm}
    \caption{
    The maximum CMB RMS noise level, $\Delta_T^{(\rm max)}$, that warrants an $\textrm{SNR (cross)} \geq 3$ detection for the optical-depth signal, for a range of ionization fraction function parameter values. 
    Here, $L_{\rm max}$ is assumed to be $6000$, and the bubble size distribution parameters are fixed to their fiducial values indicated in Tab.~\ref{tab:set_of_params}.
    The right (left) panel corresponds to a CMB-S4-like (CMB-HD-like) beam. 
    Unlike previous detection-SNR-related figures, here we set the sky fraction $f_{\rm sky} =0.4 $, matching anticipated sky coverage from combinations of CMB-S4 and CMB-HD with LSST. 
    Note the significant variation of $\Delta_T^{(\rm max)}$ values with varying $y_{\rm re}^{\rm He}$ and $\Delta_y^{\rm He}$ values. This figure is similar to Fig.~\ref{fig:SNR_contour_xe} in demonstrating that an unambiguous detection of optical depth at a certain significance can in principle be used to pose bounds on the helium-reionization parameter space.
    }
    \label{fig:DeltaT_max_contour_xe}
    \vspace*{-0.45cm}
\end{figure}

\begin{figure}[h!]
    \centering
    \includegraphics[width=0.85\textwidth]{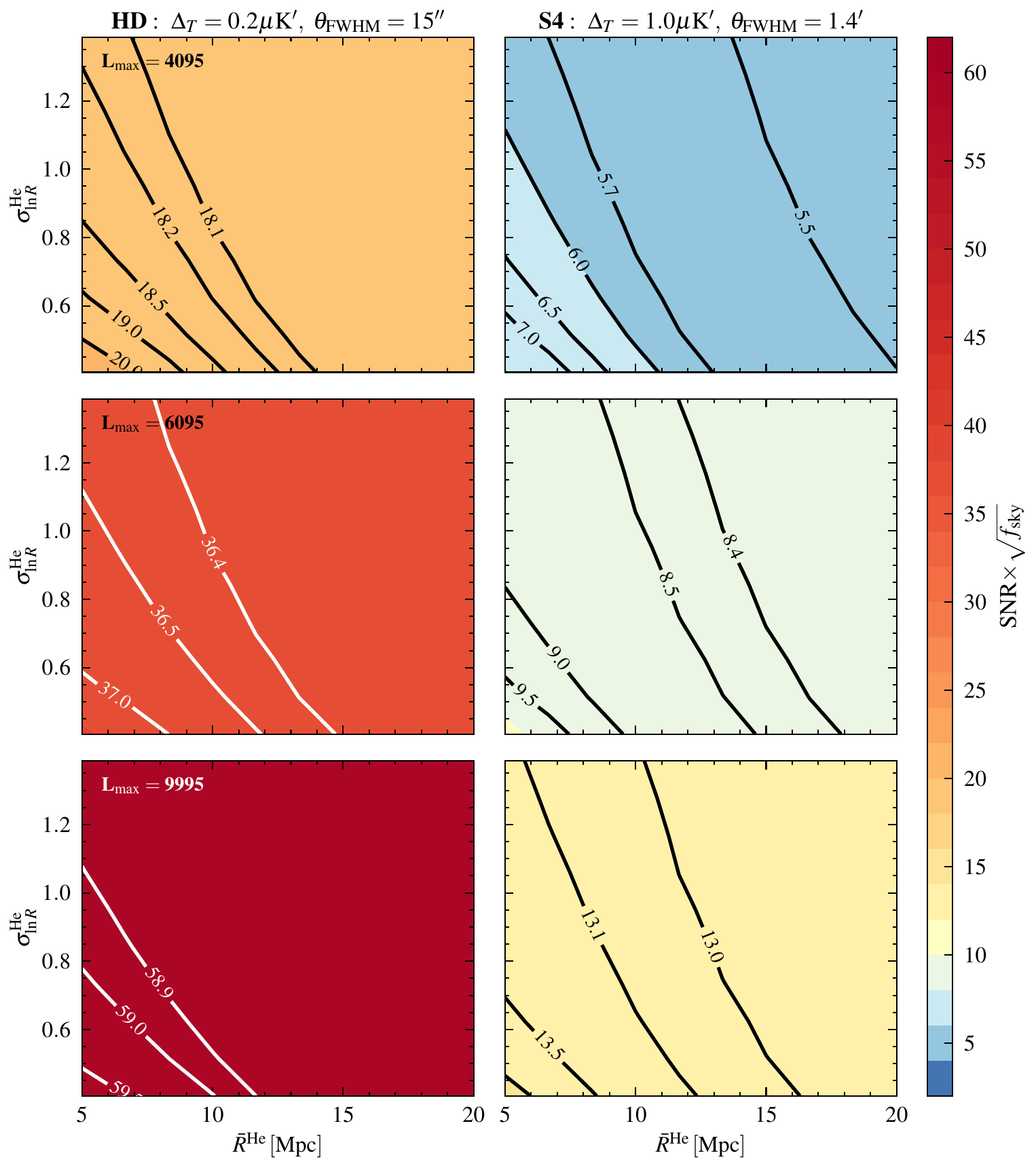} 
    \vspace*{-0.3cm}
    \caption{
    Similar to Fig.~\ref{fig:SNR_contour_xe}, except bubble size distribution parameters $\bar{R}^{\rm He}$ and $\sigma^{\rm He}_{\ln{R}}$ are varied while the ionization fraction parameters are set to their fiducial values highlighted in Tab.~\ref{tab:set_of_params}.
    Unlike $y_{\rm re}^{\rm He}$ and $\Delta_y^{\rm He}$ (cf. Fig.~\ref{fig:SNR_contour_xe}), a variation in the bubble size distribution parameters has little effect on the detection SNR.
    This is expected, as the detection SNR is dominated by the $\langle\tau g\rangle$-cross signal, which is only weakly sensitive to the bubble size distributions considered in our study.
    }
    \label{fig:SNR_contour_bubble}
    \vspace*{-0.5cm}
\end{figure}

While the results shown in Figs.~\ref{fig:SNR_1} and~\ref{fig:SNR_2} are based on the fiducial hydrogen and helium reionization parameters outlined in Tab.~\ref{tab:set_of_params},
we show the dependence of the optical-depth detection SNR on varying helium reionization parameters in Figs.~\ref{fig:SNR_contour_xe}~and~\ref{fig:DeltaT_max_contour_xe}.
Figure~\ref{fig:SNR_contour_xe} shows the detection SNR as a function varying helium reionization parameters $y_{\rm re}^{\rm He}$ and $\Delta_y^{\rm He}$.
Here, the bubble size distribution parameters are fixed at their fiducial values, and $f_{\rm sky}$ is normalized to unity.
Similar to before, the detection SNR can be scaled by $\sqrt{f_{\rm sky}}$ depending the sky coverage fraction of a given survey.
The top, middle and bottom panels correspond to $L_{\rm max}$ choices of 4095, 6095, and 9995, respectively. 
The left (right) column corresponds to CMB-HD (CMB-S4) like survey specifications.
In all cases, we find the detection SNR depends \textit{significantly} on the helium reionization parameter values, indicative of the sensitivity of the total signal to characteristics of helium reionization.
In addition to the trivial increase of SNR with increasing $L_{\rm max}$, the inclination of the fixed-SNR contours indicates that detection prospects improve with decreasing $y_{\rm re}^{\rm He}$ and increasing $\Delta_y^{\rm He}$. In other words, the detection SNR for helium reionization is higher when helium reionization occurs at a relatively later redshift over a longer duration, with highest SNR achieved for $\bar{x}_e^{\rm He}$ modelled according to the orange line in Fig.~\ref{fig:He_variation}. An increased duration corresponds to an extended period of `patchiness', leading to larger effect of the optical depth fluctuations on the integrated signal measurement. Furthermore, $\langle\tau g\rangle$ is the dominant contribution to the detection SNR (see Figs.~\ref{fig:SNR_1} and~\ref{fig:SNR_2}), with tomographic galaxy measurements improving in later redshift bins. As a result, a later epoch of reionization allows for a stronger cross-correlation between the optical-depth and the galaxy-density field leading to a higher detection SNR.

Similarly, Fig.~\ref{fig:DeltaT_max_contour_xe} shows the highest RMS CMB noise level $\Delta_T^{\rm (max)}$ resulting in an $\textrm{SNR} \geq 3$ detection for the reconstructed optical depth for a $1.4'$ ($15''$) beam on the left (right) panel. 
Note that, unlike the previous detection SNR figures, we set $f_{\rm sky}=0.4$ here.
Again, we find that varying helium reionization parameters leads to significant variations of $\Delta_T^{\rm (max)}$, suggesting a detection of the optical-depth fluctuations at a given SNR and experimental noise can in principle be translated to upper or lower limits on the helium reionization parameter space.

Figure~\ref{fig:SNR_contour_bubble} shows the same as Fig.~\ref{fig:SNR_contour_xe} except bubble size distribution parameters $\bar{R}^{\rm He}$ and $\sigma^{\rm He}_{\ln{R}}$ are varied while the ionization fraction parameters are set to their fiducial values highlighted in Tab.~\ref{tab:set_of_params}.
We find that variation in the bubble size distribution has little effect on the detection SNR compared to the previous case (cf.~Fig.~\ref{fig:SNR_contour_xe}).
This is expected, as the detection SNR is dominated by the $\tau\ \textrm{--}\ g$ cross signal, which is mainly sensitive to scales larger than that of bubble sizes considered in this study. 
We omit the version of Fig.~\ref{fig:DeltaT_max_contour_xe} where the bubble size distribution parameters are varied since, similar to before, variation of these parameters does not greatly change the results.

The detection SNR results shown in this section demonstrate the total statistical power of CMB and galaxy surveys for detecting optical-depth fluctuations from the ionized second electron in helium, including contributions from electron fluctuations during late times after the end of helium reionization.
As a result, they do not directly correspond to the detection prospects of helium reionization parameters, which we investigate in the next section with information-matrix forecasts.

\subsection{Probing helium reionization with the patchy optical depth and LSS}\label{sec:fisher_forecasts}

To assess future prospects of characterizing helium reionization via the detection of reionization model parameters from CMB and LSS (galaxies), we define an ensemble-information matrix as
\be\label{eq:fisher}
\mathcal{F}_{ik}=\sum\limits_{XYWZ}\sum\limits_{\ell=L_{\rm min}}^{L_{\rm max}}\!\!\frac{\partial C_\ell^{X Y}}{\partial {\pi_i}}\textbf{cov}^{-1} \left(\tilde{C}^{X W}_{\ell},\tilde{C}^{Y Z}_{\ell'}\right)\frac{\partial C_{\ell'}^{W {Z}}}{\partial {\pi_k}}\,,
\ee
where $XY, ZW \in \{{\tau}_\alpha {\tau}_\beta, {\tau}_\alpha g_\beta, g_\alpha g_\beta \}$ for every unique pair of redshift bins centered at $z_\alpha$ and $z_{\beta}$. 
Here, $\partial C_\ell^{X}/\partial\pi_i$ represents the derivative of the signal with respect to the parameter $\pi_i$. 
Our parameter array includes five helium reionization parameters $\{y_{\rm re}^{\rm He},\Delta_{\rm re}^{\rm He},b_{\rm He},\sigma_{\rm \ln R}^{\rm He},\bar{R}^{\rm He}\}$, as well as another five parameters characterising hydrogen reionization $\{y_{\rm re}^{\rm H},\Delta_{\rm re}^{\rm H},b_{\rm H},\sigma_{\rm \ln R}^{\rm H},\bar{R}^{\rm H}\}$ and the galaxy bias $b_g$, as defined in Sec.~\ref{sec:He_reionization_model}. 

\begin{figure}[t!]
    \centering
    \includegraphics[width=1\textwidth]{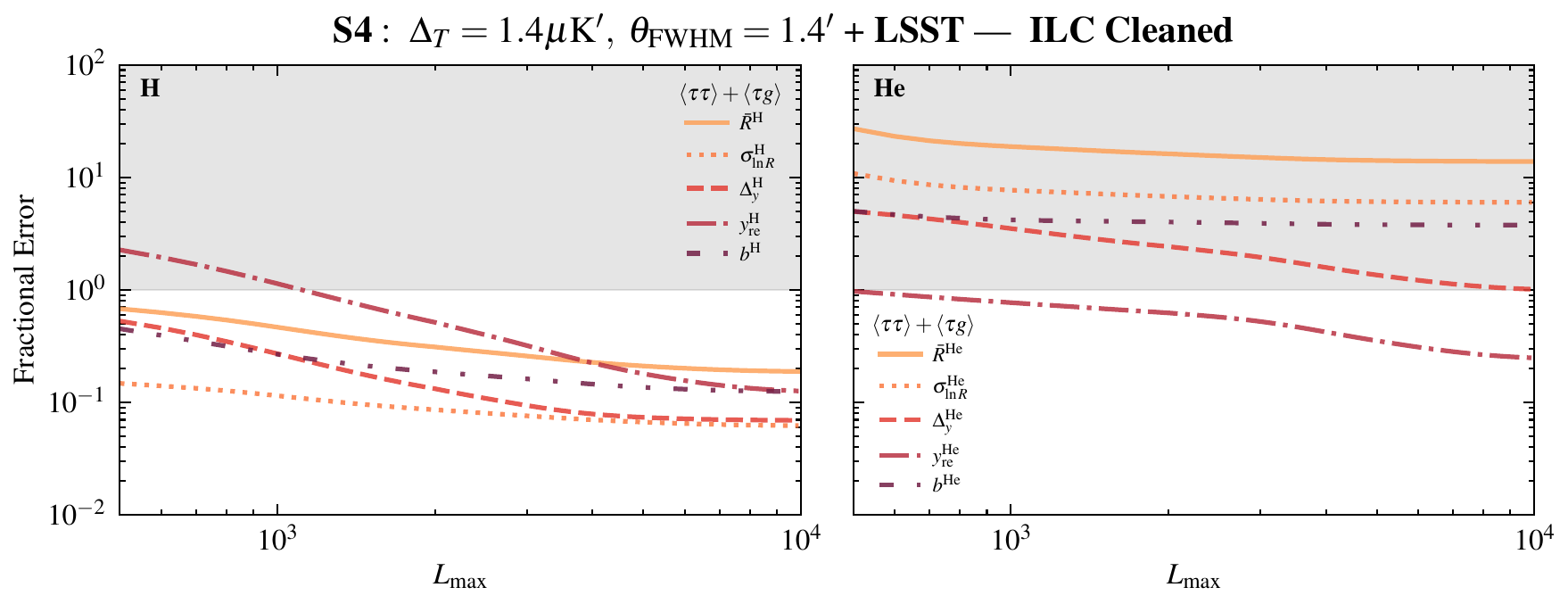} 
    \vspace*{-0.8cm}
    \caption{Fractional errors of helium (right panel) and hydrogen (left panel) reionization parameters. 
    The $y$-axes are fractional errors defined as $\Delta(\pi_i)/ \pi_i$, where $\pi=\{\bar{R}^\theta, \sigma_{\ln R}^\theta, \Delta_y^{\theta}, y_{\rm re}^\theta, b^\theta\}$ with $\theta=\{\rm He,H\}$ for helium and hydrogen reionization parameters, respectively. 
    The $x$-axes correspond to the highest multipole of the reconstructed optical depth $L_{\rm max}$ taken in the information-matrix calculation. 
    Here, we consider a CMB-S4 like survey after ILC cleaning, together with LSST. The gray-shaded region corresponds to fractional errors larger than unity, below which parameters could be detected. 
    We find that while the optical-depth reconstruction will provide high-accuracy measurements of all hydrogen-reionization-model parameters we consider, measuring helium reionization may be more difficult. 
    For our fiducial helium-reionization model choices, only the redshift of the patchy epoch of helium reionization may be measured to high precision with CMB-S4 and LSST. 
    We note, however, that these results depend on our modeling choices and if helium reionization occurs at lower redshift or lasts longer, detection prospects could be more optimistic as shown in Fig.~\ref{fig:fisher_contour_ILC_S4}.}
    \label{fig:fisher_1D_ILC_S4}
    \vspace*{-0.0cm}
\end{figure}

Our data array consists of reconstructed optical-depth fluctuations in 16 redshift bins within the range $z\in[0.2,20]$---8 bins distributed within $z\in[0.2,5.0]$ and another 8 bins within $z\in[5.0,20]$, with equal spacing in redshift. We include the galaxy field only for the 8 lower-redshift bins with otherwise identical redshift spacing. 
It is important to note that although our information matrix construction appropriately assumes that the signals at different $\ell$-modes are independent, it does not assume that the observed field is uncorrelated across different redshift bins. 
As stated earlier, we do assume that the optical depth field and the galaxy over-density field show minimal correlation across redshift bins, the non-zero correlation is purely sourced by the \textit{unbiased} ${\tau}$-reconstruction noise ${{N}_{\ell,\alpha\beta}^{\tau\tau}}$ derived in Sec.~\ref{sec:tau_estimator}. 

With the previously defined experiment configurations, and the above assumptions on redshift-binning and signal correlations, we finally construct an 11-dimensional information matrix via Eq.~\eqref{eq:fisher} to estimate our constraining power on the fiducial reionization parameters (see Tab.~\ref{tab:set_of_params}). 
Note that the inclusion of both H and He reionization signals in this single information matrix allows for \textit{forecasts that account for possible parameter degeneracies between the two epochs}. 
Furthermore, as we will show below, an exploration of the helium reionization parameter space under this information-matrix formalism will allow for predictions on the impact of the helium signal on our ability to constrain standard hydrogen reionization parameters. Note that we only perform the following analysis with the inclusion of the galaxy cross-correlation i.e., we do not provide parameter constraints using the optical depth field alone because the unbiased reconstruction noise ${{N}_{\ell,\alpha\beta}^{\tau\tau}}$ is too large to constrain all 10 reionization parameters using the $C_{\ell}^{\tau\tau}$ signal alone.

\begin{figure}[t!]
    \centering
    \includegraphics[width=1\textwidth]{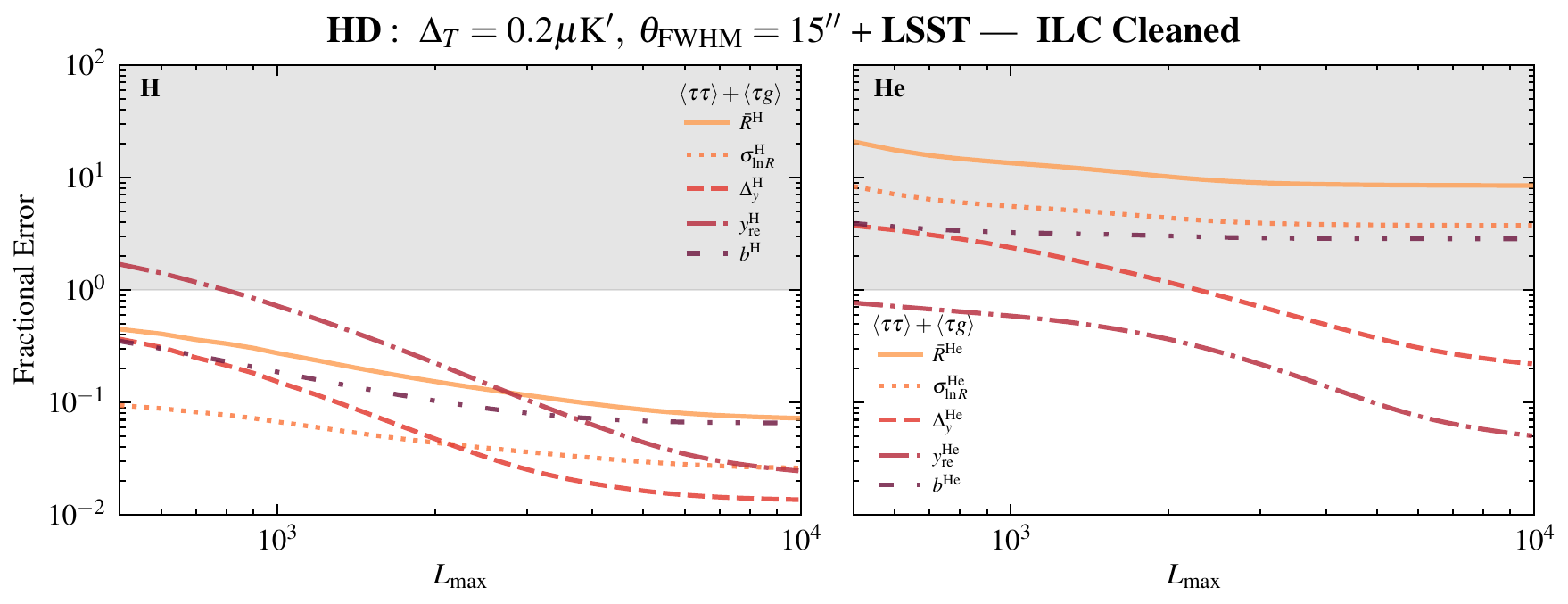} 
    \vspace*{-0.7cm}
    \caption{
    Similar to Fig.~\ref{fig:fisher_1D_ILC_S4}, but in this case, we consider a CMB-HD-like experiment with ILC cleaning. 
    We find that, for sufficiently high values of $L_{\rm max}$,
    both of the helium reionization parameters characterising the redshift and duration of the patchy epoch can be measured to high accuracy. 
    We also note that the measurement accuracy for hydrogen reionization parameters increases significantly compared to CMB-S4, suggesting future CMB surveys such as CMB-HD, when combined with LSST, will play a significant role in characterizing both hydrogen and helium reionization to high precision.
    }
    \vspace*{-0.45cm}
    \label{fig:fisher_1D_ILC_HD}
\end{figure}

The fractional errors on the fiducial helium and hydrogen reionization parameters can be found in Figs.~\ref{fig:fisher_1D_ILC_S4}~and~\ref{fig:fisher_1D_ILC_HD}. 
Figure~\ref{fig:fisher_1D_ILC_S4} shows the fractional errors from combination of CMB-S4 and LSST, whereas Fig.~\ref{fig:fisher_1D_ILC_HD} displays the same results, this time assuming an experimental configuration corresponding to CMB-HD and LSST. 
Both sets of results assume that the optical-depth signal is obtained from the CMB using ILC cleaning and the constraints are displayed as a function of the maximum reconstructed multipole $L_{\rm max}$ in Eq.~\eqref{eq:fisher}.
In each case, the right (left) panel corresponds to fractional errors on helium (hydrogen) reionization parameters.

\begin{figure}[t!]
    \centering
    \includegraphics[width=0.85\textwidth]{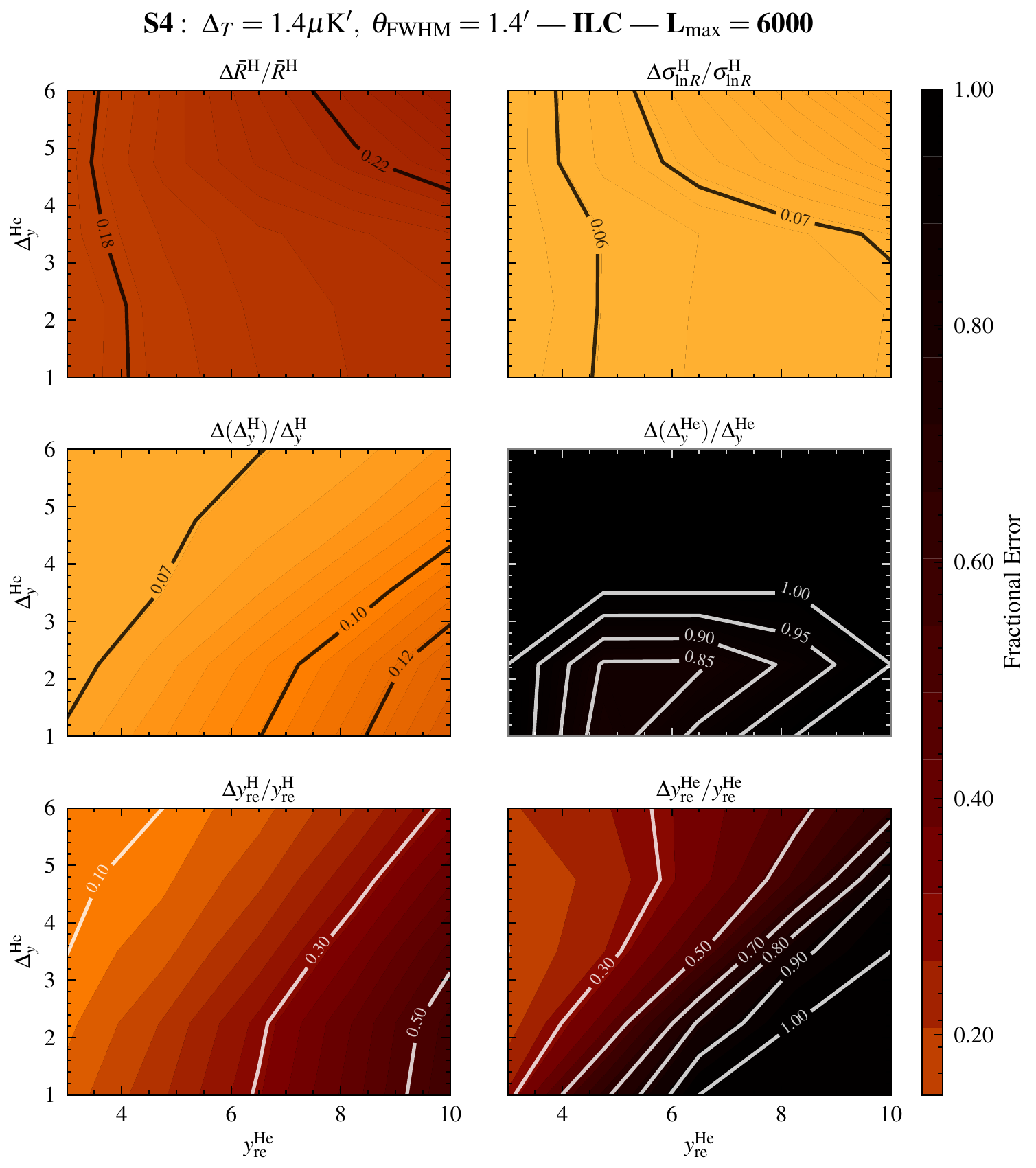} 
    \vspace*{-0.5cm}
    \caption{Dependence of fractional errors on varying values of helium reionization parameters $y_{\rm re}^{\rm He}$ and $\Delta_y^{\rm He}$, similar to Fig.~\ref{fig:SNR_contour_xe}, for a CMB-S4 like survey together with LSST. 
    Here, we set $L_{\rm max} = 6000$ and assume the CMB spectra to be ILC cleaned.
    The lower two panels in the right column correspond to fractional errors on helium ionization-fraction parameters. The remainder of the panels correspond to parameters modeling hydrogen reionization. We find that for all parameters, fractional errors depend on the fiducial choice of $y_{\rm re}^{\rm He}$ and $\Delta_y^{\rm He}$. 
    This is particularly important for probing hydrogen reionization, for which the constraints on parameters such as $\Delta_y^{\rm H}$ and $y_{\rm re}^{\rm H}$ vary by up to a factor of $\mathcal{O}(5)$ within the range of the parameter space we consider here, with errors becoming larger with shorter duration  of helium reionization (lower $\Delta_y^{\rm He}$) and with higher values of $y_{\rm re}^{\rm He}$.
    This is due to the degeneracy between helium and hydrogen reionization parameters in case helium reionization occurs before hydrogen reionization ends completely. Our results suggest that reionization of helium could, in principle, play a limiting role on the non-ambiguous characterisation of hydrogen reionization from optical-depth reconstruction.
    }
    \label{fig:fisher_contour_ILC_S4}
    \vspace*{-0.5cm}
\end{figure}

\begin{figure}[h!]
    \centering
    \includegraphics[width=0.85\textwidth]{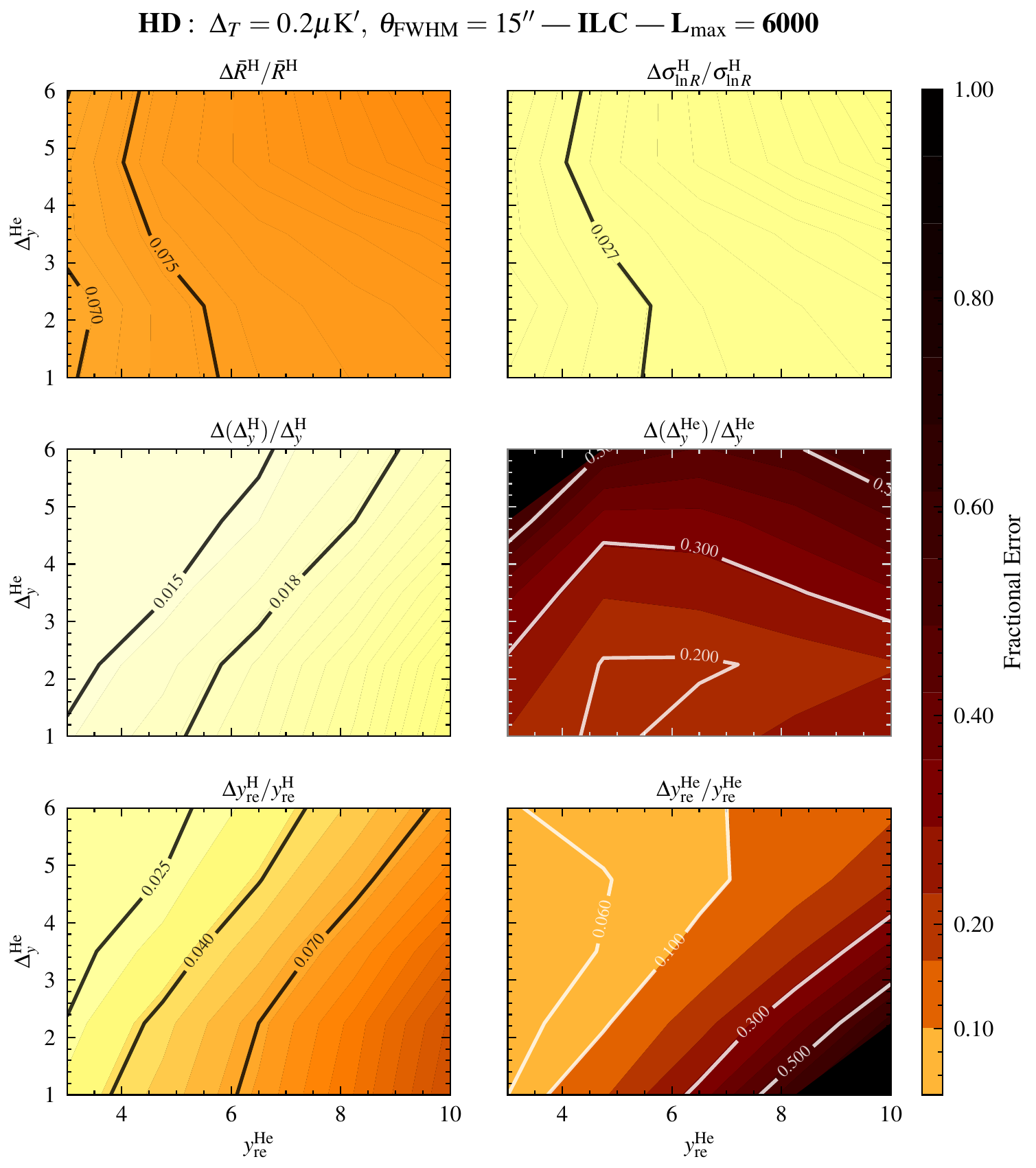} 
    \vspace*{-0.5cm}
    \caption{Similar to Fig.~\ref{fig:fisher_contour_ILC_S4}, except the results are for a CMB-HD-like survey together with LSST.
    As seen, the measurability of both the He ionization-fraction parameters improve compared to case with CMB-S4, and both $y_{\rm re}^{\rm He}$ and $\Delta_y^{\rm He}$ can be constrained across most of the grid.
    This is followed with an additional improvement to the measurement of H reionization parameters compared to the previous case.
    These results indicate that the upcoming CMB and galaxy surveys will have the statistical power to probe the midpoint of helium reionization in redshift and the duration of the patchy epoch (parametrized by $y_{\rm re}^{\rm He}$ and $\Delta_y^{\rm He}$, respectively) to a sufficient precision via optical-depth reconstruction cross-correlated with the statistically powerful large-scale structure data sets.
    }
    \label{fig:fisher_contour_ILC_HD}
    \vspace*{-0.5cm}
\end{figure}

\begin{figure}[h!]
    \centering
    \includegraphics[width=1.0\textwidth]{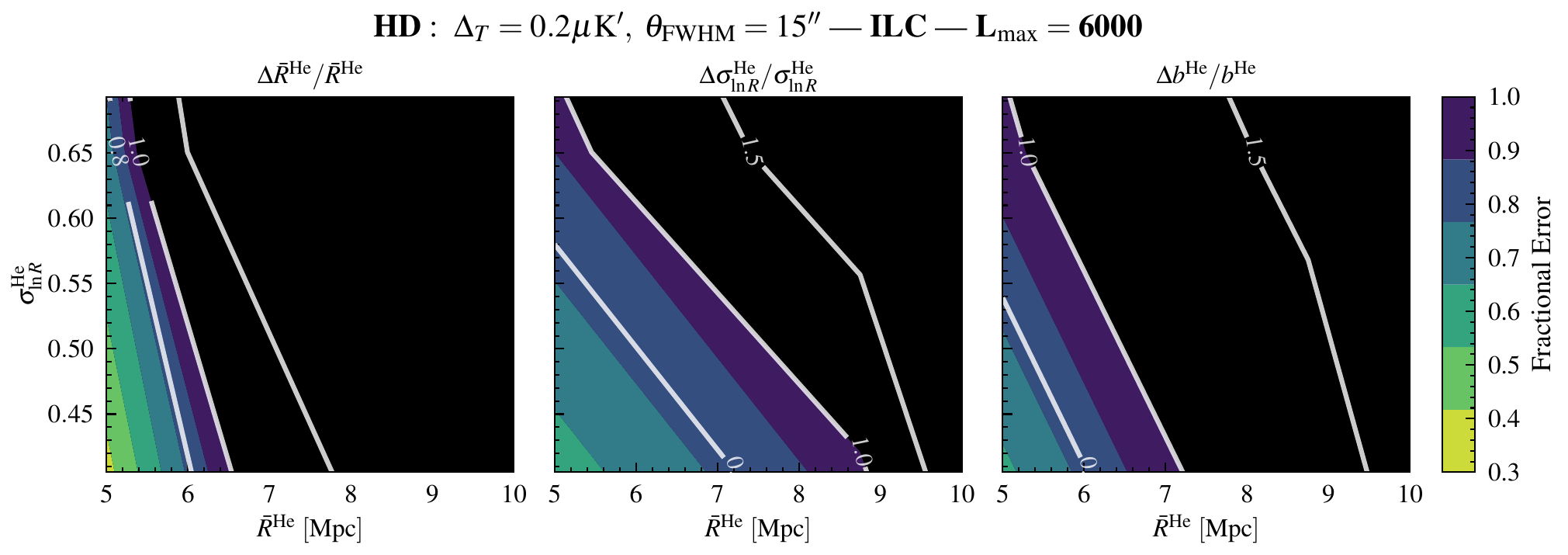} 
    \vspace*{-0.5cm}
    \caption{
    Dependence of the measurability of He bubble size distribution parameters, $\bar{R}^{\rm He}$ and $\sigma^{\rm He}_{\ln{R}}$, and the Helium bubble bias $b^{\rm He}$ on varying $\bar{R}^{\rm He}$ and $\sigma^{\rm He}_{\ln{R}}$ for CMB-HD-like survey specifications after ILC cleaning together with LSST.
    For simplicity, we only show the results for $L_{\rm max} = 6000$.
    Colored regions represent where the fractional errors $\Delta \pi_i/\pi_i <1$, while the regions where $\Delta \pi_i/\pi_i \geq 1$ are colored in black.
    Results indicate $\bar{R}^{\rm He}$, $\sigma^{\rm He}_{\ln{R}}$, and $b^{\rm He}$ can be constrained if the bubble sizes are small enough ($\bar{R}^{\rm He} \lesssim 6\textrm{--}9$ Mpc) and the distribution width is low $\sigma^{\rm He}_{\ln{R}} \lesssim \ln{1.8}$. 
    Note that the range of $\bar{R}^{\rm He}$ and $\sigma_{\ln{R}}^{\rm He}$ displayed here is narrower than that in previous plots to enhance the visibility of the measurable parameter space.
    }
    \label{fig:fisher_contour_bub_ILC_HD}
    \vspace*{-0.5cm}
\end{figure}

In agreement with forecasts in the literature~\citep[see, e.g.,][]{Jain:2023xdy,Bianchini:2022wte,Guzman:2021ygf,Guzman:2021nfk,Paul:2020fio,Feng:2018eal,Roy:2018gcv,Meyers:2017rtf,Fidler:2017irr}, we find that optical depth reconstruction, via both the assumed CMB experiment specifications, will allow for high-precision measurements of all hydrogen reionization parameters, further demonstrating the potential value of this program in the near future.
In contrast, the information-matrix results indicate that characterizing the helium epoch will be more cumbersome. 
For the combination of CMB-S4 and LSST, we find that fractional errors reach values below unity only for $y_{\rm re}^{\rm He}$, indicating that only the redshift of the patchy epoch (midpoint) of helium reionization may be constrained.

Nonetheless, for the combination of CMB-HD and LSST, we see an improvement, with below-unity fractional errors achieved for both $y_{\rm re}^{\rm He}$ and $\Delta_{\rm re}^{\rm He}$ at sufficiently high $L_{\rm max}$. This is followed with an additional improvement to the measurement of hydrogen reionization parameters compared to the previous case.
This implies that upcoming CMB and galaxy surveys will have the statistical power to constrain the midpoint of helium reionization in redshift and the duration of the patchy epoch (parametrized by parametrized by $y_{\rm re}^{\rm He}$ and $\Delta_{\rm re}^{\rm He}$, respectively) to a sufficient precision via optical-depth reconstruction cross-correlated with the statistically powerful galaxy survey data sets. 

Advances in ILC-cleaning methods can, in principle, further improve these constraints, as shown in Figs.~\ref{fig:fisher_1D_BB_S4}~and~\ref{fig:fisher_1D_BB_HD}, resulting in, for example, multiple helium reionization parameters being detected for sufficiently large $L_{\rm max}$ values even with CMB-S4. 
In all the experiment configurations considered, under the fiducial parameter values (Tab.~\ref{tab:set_of_params}), the helium bubble-size-distribution parameters $\{\bar{R}^{\rm He}, \sigma_{\ln R}^{\rm He}\}$ and bubble bias $b^{\rm He}$ remain unmeasurable ($\Delta \pi_i/\pi_i > 1.0$). 
This is because the constraining power during the epoch of He reionization is primarily sourced by the cross-correlation $\langle\tau g\rangle$. 
Our current model for this correlation only accounts for galaxies as a \textit{large-scale} tracer of bubbles i.e.; it only accounts for correlations of the 2-bubble term of galaxy distributions and optical depth fluctuations, where the bubble bias is degenerate with fluctuations in the bubble size distribution [see Eq.~\ref{eq:powespec_cross}]. Therefore, relying solely on $P_{\Delta_e\Delta_e}^{1b, {\rm He}}$ for constraints on the bubble parameters does not result in enough power to characterize the morphology of ionized regions during this epoch.

Similarly to our SNR analysis, we once again vary the assumed set of fiducial He reionization parameters to assess the impact of the morphology of this epoch on the measurement prospects of not only the helium parameters but also their hydrogen counterparts. 
We first consider the resulting parameter constraints under variations of the redshift dependence/evolution of the helium reionization epoch. We run the information-matrix formalism, varying the assumed shape of $\bar{x}_e^{\rm He}(z)$ with $3.0 \leq y_{\rm re}^{\rm He} \leq 10.0$ and $1.0 \leq \Delta_y^{\rm He} \leq 6.0$ (as summarized in Tab.~\ref{tab:set_of_params}), holding all other parameters fixed at their fiducial values.\footnote{\label{ftn:He_var_model_extrm}Note that the model-extremes of $\bar{x}_e^{\rm He}$ [$P(R)^{\rm He}$] considered in this analysis correspond to the two variations plotted in the top [bottom] panel of Fig.~\ref{fig:He_variation}, over the assumed fiducial model in red.} 
The results of these parameter variations, assuming ILC-cleaned CMB observations, are summarized in
Figs.~\ref{fig:fisher_contour_ILC_S4}~and~\ref{fig:fisher_contour_ILC_HD}, with the former (latter) displaying the relevant results for the experiment configuration corresponding to CMB-S4 (CMB-HD) and LSST. 
In each of the figures, the top row presents the effects of varying $y_{\rm re}^{\rm He}$ and $\Delta_y^{\rm He}$ on the fractional error of the hydrogen bubble parameters $\bar{R}^{\rm H}$ and $\sigma_{\ln R}^{\rm H}$. 
Similarly, the middle (bottom) panel displays the effects of this variation on the duration (mean-reionization redshift) of each of the epochs, characterized by $\Delta_y^{\rm H}$ and $\Delta_y^{\rm He}$ ($y_{\rm re}^{\rm H}$ and $y_{\rm re}^{\rm He}$). 
We do not display the fractional error contour plots for helium bubble parameters $\bar{R}^{\rm He}$, $\sigma_{\ln R}^{\rm He}$, and the bubble bias $b^{\rm He}$, since these parameters are not measurable ($\Delta \pi_i / \pi_i > 1.0$) at any point in the grid. 

Across both the considered experiment configurations, the results indicate that if helium reionization occurs at lower redshifts (lower $y_{\rm re}^{\rm He}$ values), or lasts a longer time (larger $\Delta_y^{\rm He}$ values) fractional errors on $y_{\rm re}^{\rm He}$ improve significantly, reducing by around an order of magnitude within the span of the parameter space we consider. 
The behaviour of fractional errors in $\Delta_y^{\rm He}$ is more complicated, with information-matrix results indicating improved constraints for mid-range $y_{\rm re} \sim 5.0$ with shorter duration of transition (lower $\Delta_y^{\rm He}$).

Furthermore, it is clear that the measurability of some hydrogen reionization parameters is also visibly impacted, up to a factor of $\mathcal{O}(5)$, by variations in $y_{\rm re}^{\rm He}$ and $\Delta_y^{\rm He}$. 
The parameters characterizing $\bar{x}_e^{\rm H}$ see decreasing fractional errors with increasing duration of the helium epoch and decreasing mean redshift, whereas the hydrogen bubble parameters see minimal change across the grid. 
This behaviour can be attributed to an enhanced degeneracy between the helium and hydrogen reionization parameters if helium reionization occurs at higher redshifts, where the cross-correlation $\langle\tau g\rangle \rightarrow 0$\,, before hydrogen reionization comes to a complete end.  
Therefore, Figs.~\ref{fig:fisher_contour_ILC_S4}~and~\ref{fig:fisher_contour_ILC_HD} suggest that characterization of hydrogen reionization from the reconstructed optical depth can, in principle, be biased by helium reionization.

Our analysis indicates that varying the He bubble size distribution parameters $\bar{R}^{\rm He}$ and $\sigma^{\rm He}_{\ln{R}}$ have minimal $(\lesssim 10 \%)$ impact on the measurability of hydrogen reionization parameters. Similarly, the measurability of the helium ionization fraction parameters $y_{\rm re}^{\rm He}$ and $\Delta_y^{\rm He}$ also do not depend heavily on varying He bubble parameters.
Nonetheless, we find that varying the He bubble size distribution have significant impact on the measurability of He bubble parameters $\bar{R}^{\rm He}$ and $\sigma^{\rm He}_{\ln{R}}$ as well as the He bubble bias $b^{\rm He}$.

To quantify the extent of this impact, we vary helium bubble distribution $P(R)^{\rm He}$, with $5.0\ \textrm{Mpc} \leq \bar{R}^{\rm He} \leq 20.0\ \textrm{Mpc}$ and $\ln 1.5 \leq \sigma_{\ln R}^{\rm He} \leq \ln 4.0$ (as summarized in Tab.~\ref{tab:set_of_params}), holding all other parameters fixed at their fiducial values\footref{ftn:He_var_model_extrm}. 
Figure~\ref{fig:fisher_contour_bub_ILC_HD} shows the fractional errors for $\bar{R}^{\rm He}$ (left), $\sigma_{\ln R}^{\rm H}$ (middle) and helium bubble bias $b^{\rm He}$ (right).
Here, we show the results for $L_{\rm max} = 6000$ and assume that the CMB spectra are ILC cleaned.

Despite the concentrated effects of bubble-parameter variations, all three contour plots agree on an important take-away---although initial information matrix results on the physically motivated, fiducial parameter space may indicate that some helium parameters may never be measurable, results can change starkly with changes in the assumed bubble distribution. Specifically, the three previously unmeasurable ($\Delta \pi_i / \pi_i > 1.0$) parameters  $\bar{R}^{\rm He}$, $\sigma_{\ln R}^{\rm He}$, and $b^{\rm He}$ can be measured ($\Delta \pi_i / \pi_i < 1.0$) if the regions of ionized helium are smaller ($\bar{R}^{\rm He} \lesssim 6$ Mpc). The conditions are slightly looser for $\sigma_{\ln R}^{\rm He}$ in specific, which remains measurable up to $\bar{R}^{\rm He} \lesssim 9$ Mpc, as long as the distribution of bubbles is sharply peaked at the mean value. This behaviour can be attributed to the fact that the ionized bubble distribution strongly effects the small-scale 1-bubble term of $P^{\rm He}_{\Delta_e\Delta_e}$ appearing in our model for $C_{\ell}^{\tau\tau}$. Smaller bubble size boosts the power in the $P^{1b, {\rm He}}_{\Delta_e\Delta_e}$ making the model more sensitive to changes in the $P(R)^{\rm He}$. The effect of smaller characteristic bubble sizes can be seen, for example, in the middle panel (orange line) of Fig.~\ref{fig:varied_He_and_cross}. This manifests in our information-matrix analysis in the form of increasingly tight constraints on $\bar{R}^{\rm He}$, $\sigma_{\ln R}^{\rm He}$, and $b^{\rm He}$.

\section{Discussion} \label{sec:discussion}

Over the course of cosmic history, the intergalactic medium experienced two pivotal transitions: the reionization of hydrogen and helium. 
While hydrogen reionization, primarily driven by photons from hot stars, has been extensively studied, helium reionization, occurring at lower redshifts around $z\sim3$, has garnered less attention. 
Helium reionization is anticipated to be anisotropic, like that of hydrogen, characterized by the formation and growth of ionized bubbles around luminous sources~\citep{Wyithe:2002qu,Madau:2015cga,McQuinn:2012bq,Worseck:2014gva,Furlanetto:2007mg,Worseck:2011qk,Sokasian:2001xh,Compostella:2013zya,Oh:2000sg,Furlanetto:2007gn,Furlanetto:2008qy,Dixon:2009xa,LaPlante:2016bzu,Caleb:2019apf,Linder:2020aru,Meiksin:2011bq,Compostella:2014joa,Eide:2020xyi,UptonSanderbeck:2020zla,Bhattacharya:2020rtf,Villasenor:2021ksg,Meiksin:2010rv,Gotberg:2019uhh,LaPlante:2015rea,Syphers:2011uw,Dixon:2013gea,LaPlante:2017xzz,Lau:2020chu,Hotinli:2022jna,Hotinli:2022jnt}.
However, the reionization of helium is driven by energetic photons from active galactic nuclei and quasars, since the second electron in helium requires $\sim 54{\rm eV}$ of energy to be ionized. 
As a result, the morphology of helium reionization can provide valuable insight into quasar abundance, clustering, luminosities, variability, and lifetimes, and the early formation of supermassive black holes~\cite{2012ApJ...755..169M,2013ApJ...773...14R,2013ApJ...768..105M,McGreer:2017myu,2022ApJ...928..172P, Shen:2014rka,Hopkins:2006vv,2017ApJ...847...81S}.

In this study, we have evaluated the potential of forthcoming surveys to probe signatures of helium reionization through the reconstruction of the optical-depth field $\tau$. 
We repurpose existing statistical tools of optical depth reconstruction, previously developed for probing hydrogen reionization in Refs.~\citep{Dore:2007bz, 2007ApJ...663L...1H, Dvorkin:2008tf, Dvorkin:2009ah, Feng:2018eal}, and combine it with the statistical power of galaxy surveys, taking advantage of the late-time occurrence of helium reionization to forecast that the characterization of the morphology of this epoch is achievable in the near future.

We characterize the epoch of helium reionization by modeling the evolution of the mean ionization fraction of helium $\bar{x}_e^{\rm He}(z)$ and the size distribution of `bubble'-like regions of ionized helium. The modeling choices are an extension of the functional forms previously used to characterize the epoch of hydrogen reionization (as seen in Refs.~\citep{Dvorkin:2008tf, Dvorkin:2009ah}), with an updated parameter space to account for the later occurrence of the helium epoch and the difference in the ionizing sources. Based on this morphology, we model the expected optical-depth-fluctuation power spectrum $C_{\ell}^{\tau\tau}$, critically accounting for \textit{both} helium and hydrogen reionization (as an extension of the model presented in Ref.~\citep{Dvorkin:2008tf}). 
For the first time, we also present a model for the cross-correlation of the $\tau$-field with the galaxy density field that accounts for both of these separate epochs. 
Additionally, we present a derivation for the expected noise in the reconstruction of the optical-depth field, assuming that optical-depth measurements are derived from their effects on the CMB temperature and polarization maps.

With the noise and signal defined, we present forecasts on our ability to measure the effects of optical-depth fluctuations sourced by helium reionization in the form of measurement SNR. 
To estimate how well the conjunction of future CMB experiments and galaxy surveys may be able to characterize the reionization morphology---including the time, duration, and patchiness of the epoch---we perform a comprehensive information-matrix analysis that incorporates the parameter space characterizing both the helium and hydrogen reionization.
Thus, we also account for any possible modeling degeneracies that may arise due to the similarities in the signal from the two separate epochs. 

In our forecasts, we primarily consider two sets of experimental configurations, both of which approximate the specifications of either CMB-S4~\citep{CMB-S4:2016ple} or CMB-HD~\citep{CMB-HD:2022bsz}.
In each case, we assume that the CMB-reconstructed optical depth is cross-correlated with the tomographic galaxy survey data obtained from the LSST~\citep{LSSTDarkEnergyScience:2018jkl}. 
We perform forecasts for varying choices of $L_{\rm max}$ (maximum multipole of $\tau$-reconstruction) and different assumptions on the CMB foreground.
The conservative forecasts assume that the optical-depth field is obtained from the CMB using ILC cleaning, while the more optimistic ones assume that the CMB dataset is reduced only to its black-body contributions. Note also that we omitted the forecasting for MegaMapper~\citep{Schlegel:2019eqc}, which is planned to be the spectroscopic follow-up to LSST. Such a survey will measure galaxy redshifts to high precision, allowing the use of the full three-dimensional galaxy distribution much more thoroughly.
We leave the inclusion of this survey in forecasting to future work.

Through this analysis, we find that the characterization of the epoch of helium reionization is primarily driven by the power in the cross-correlation $\langle \tau g\rangle$, irrespective of the assumed experiment configuration. 
In specific, while combination of CMB-S4 and LSST will have sufficient statistical power to detect optical-depth fluctuations sourced by the epoch of helium reionization, connecting these detections to an unambiguous measurement of helium reionization parameters may be more difficult. 
Nevertheless, our information-matrix forecasts (on fiducial parameters) indicate that this experiment configuration will likely be able to constrain the mid-point/mean-redshift of reionization, with constraints on the duration being achievable only for more optimistic assumptions on the cleaning of the CMB maps and higher $L_{\rm max}$ of $\tau$-reconstruction. 

On the other hand, forecasts on the experiment configuration corresponding to CMB-HD and LSST indicate significant improvements, with the detection SNR for $\tau$-fluctuations sourced by helium increasing above $\mathcal{O}(10)$ for reasonable choices of foreground cleaning and $L_{\rm max}$. 
This improvement is also manifested in the parameter measurements forecasted via the information-matrix formalism. 
Our analysis suggests that the combination of CMB-HD and LSST will be able to probe the redshift evolution of helium reionization to high fidelity, pinning down the time and duration of the epoch. In both the considered experiment configurations, unfortunately, the patchy bubble parameters and bubble bias remain elusive under the fiducial assumptions on bubble size distributions.

Considering the limited constraints on the epoch of helium reionization and the subsequent flexibility in modeling, we strive to accommodate the effects of our chosen fiducial parameters. 
This is achieved by conducting (SNR and information-matrix) forecasts over a broad parameter space, thereby encompassing a diverse range of morphologies characterizing this epoch.
We vary the redshift evolution (both time and duration) of reionization and bubble size distribution to explore the parameter space where higher measurement SNRs and improved parameter constraints are likely.
Our analysis indicates that, for both experimental configurations, the measurement SNR of $\tau$-fluctuations sourced by helium reionization improves significantly when the reionization happens later, lasts longer, and the ionized regions are smaller.
In fact, for both experiment configurations, fractional errors on the mean redshift of reionization see an improvement by up to an order of magnitude as the epoch of reionization is pushed to lower redshifts and longer durations. 
Furthermore, we find that the probing the bubble size distribution becomes accessible for CMB-HD and LSST when the average size of ionized regions is small (radius of $\lesssim 7$ Mpc).

Moreover, since the modeling of the signal and the resulting information-matrix formalism accounts for both the hydrogen- and helium-reionization parameter space, we explore the effects of different helium reionization morphologies on our ability to characterize its hydrogen counterpart. 
Across both the experiment configurations, the fractional errors in the hydrogen reionization parameter space are affected the most by the changes in the redshift evolution of the helium reionization epoch.
In fact, fractional errors from our information-matrix analyses indicate that the characterization of the hydrogen epoch is easiest (and likely least biased) when the helium reionization epoch occurs at later times for longer durations. 
This behavior can be attributed to the alleviation of degeneracies across the two similarly modelled epochs, in models where helium reionization begins after hydrogen reionization is over.
Furthermore, the galaxy dataset can play a stronger role in alleviating this degeneracy if the patchiness of helium reionization occurs at lower redshifts (where the galaxies are more abundant) and lasts for a longer time (longer duration of $\langle\tau g\rangle$ correlation sourced by helium reionization). 
As a result, this work suggests that the inclusion of helium reionization in future measurements of the hydrogen reionization epoch, especially when using line-of-sight integrated signals, will be integral to ensure unbiased results on the hydrogen parameter space.

Furthermore, note that on large scales, where electron fluctuations are linear and follow dark matter (up to a bias), we model the 2-halo contribution from galaxy-hydrogen correlation to the optical-depth signal with a hydrogen bias and marginalize over this parameter. 
This approach circumvents the need to template and/or remove this contribution from our maps in practice. 
On small scales, however, the cross-correlation of free electrons (specifically those from hydrogen reionization) and galaxies remains uncertain especially at early redshifts. 
As we discussed, our analysis leaves the maximum scales probed as a free parameter and shows results for a range of scales, acknowledging the importance of this uncertainty. Nevertheless, we note that an $L_{\rm max}$ of 5000 corresponds to $k_{\rm max} \sim L_{\rm max}/\chi \lesssim 1 {\rm{Mpc}}^{-1}$ at a redshift of 2 [with $k_{\rm max}(z)$ only decreasing at higher $z$].
This means that the forecasted parameter errors in the one-dimensional information-matrix plots (see, for example, Fig.~12), for $L_{\rm max} \lesssim 5000$, are reliant on scales ($k \lesssim 1 {\rm{Mpc}}^{-1}$ for $2 \leq z \leq 5$) that are only mildly in the non-linear regime. In other words, for the lower values of $L_{\rm max}$ displayed, the errors will not be significantly impacted by uncertainties in the non-linear modeling of the late time small-scale electron distribution. 
In contrast, it is important to note that effects such as baryonic feedback introduce uncertainty in the electron distributions even on relatively larger scales ($3000<\ell<5000$).
With the current state-of-the-art simulations predicting distributions that vary up to a factor of a two at cosmic noon ($z\gtrsim2$), these modeling uncertainties may impact the forecasted errors.
However, additional measurements from other probes (such as the Sunyaev-Zel'dovich effects) at high redshifts will alleviate these issues.
We will address these points in an upcoming work, incorporating realistic priors and prospects of external measurements of the 1-halo (ionized) electron-galaxy cross-correlation.

Finally, it is vital for us to address the fact that the results presented in this paper are dependent on our simplified model of helium reionization. 
This choice was made not only to mirror existing forecasts on probing hydrogen reionization with similar techniques~\citep[e.g.,][]{Jain:2023xdy,Bianchini:2022wte,Guzman:2021ygf,Guzman:2021nfk,Paul:2020fio,Feng:2018eal,Roy:2018gcv,Meyers:2017rtf,Fidler:2017irr}, but also to model the epoch of helium reionization without including too many astrophysical priors.
In practice, both the bubble-size distribution and its mean will have redshift dependence set by the abundance of quasars (or active galactic nuclei) and their efficiency in ionizing their environment~\citep[e.g.,][]{LaPlante:2016bzu}. 
Given a distribution of sources, the mean ionization fraction can, in principle, be predicted from the typical expansion speed of ionizing bubbles. This speed may vary based on the intergalactic environment, suggesting that it is not strictly a free parameter.
Such improvements in helium (and hydrogen) reionization model building can lead to different SNR values as suggested by our analysis varying the fiducial model parameters. Furthermore, given the similarity of the modeling of the two epochs, the forecasts presented here are heavily impacted by parameter degeneracies across the two, i.e., forecasts might improve if we model the epoch of helium reionization using more priors from astrophysics, specifically given that this epoch is visibly accessible. 
We leave improvements to the modeling of helium reionization to our upcoming works on the subject. Our choices here provide a general assessment on the statistical power accessible with CMB and galaxy surveys and will guide these future efforts.

\acknowledgments
We are grateful to Simone Ferraro, Gil Holder, Amalia Madden, and Kendrick Smith for useful discussions and collaboration.
M.\c{C}., N.A.K, and M.K.~acknowledge support from the John Templeton Foundation through Grant No.~62840.
M.\c{C}.~is also supported by NSF Grants No.~AST-2006538, PHY-2207502, PHY-090003 and PHY-20043, NASA Grants No.~20-LPS20-0011 and 21-ATP21-0010, and by Johns Hopkins University through the Rowland Research Fellowship. 
M.K.~was supported by NSF Grant No.\ 2112699 and the Simons Foundation.
S.C.H.~was supported by the P.~J.~E.~Peebles Fellowship at Perimeter Institute for Theoretical Physics and the Horizon Fellowship from Johns Hopkins University. This research was supported in part by Perimeter Institute for Theoretical Physics. Research at Perimeter Institute is supported by the Government of Canada through the Department of Innovation, Science and Economic Development Canada and by the Province of Ontario through the Ministry of Research, Innovation and Science. This work was in part carried out at the Advanced Research Computing at Hopkins (ARCH) core facility (rockfish.jhu.edu), which is supported by the NSF Grant No.~OAC1920103. This work was performed in part at Aspen Center for Physics, which is supported by NSF Grant No.~PHY-2210452.

\appendix

\section{Binned optical-depth spectra}\label{sec:binned_optical_depth}

The optical depth from a range of redshifts within $z\in [z_{\rm min}^{\alpha},z_{\rm max}^{\alpha}]$ in the sky satisfies 
\be
\tau_{\theta}^{\alpha}(\nhat, z)=\sigma_T n_{p,0}\int_{z^\alpha_{\rm min}}^{z^\alpha_{\rm max}}\frac{\dd z'(1+z')^2}{H(z')}x_e^{\theta}(\nhat,z')\,,
\ee
where $\alpha$ indicates the redshift bin, and the cross-power spectra can be defined as
\be\label{eq:cross_power_covariance}
\begin{split}
C_{\ell}^{\tau_\alpha \tau_\beta} = &\sigma_T^2 n_{\rm p,0}^2 \int_{\chi_{\rm min}^{\alpha}}^{\chi_{\rm max}^{\alpha}}\!\!\dd\chi\!\!\int_{\chi_{\rm min}^{\beta}}^{\chi_{\rm max}^{\beta}}\dd\chi'\int \dd k\,k^2 
(2/\pi) \frac{j_\ell(k\chi)}{a(\chi)^2} \frac{j_{\ell}(k\chi')}{a(\chi')^2}\sqrt{P^{\rm H}_{\Delta_e\Delta_e}(\chi) P^{\rm H}_{\Delta_e\Delta_e}(\chi')}\\
&+\sigma_T^2 f_{\rm He}^2 n_{\rm p,0}^2 \int_{\chi_{\rm min}^{\alpha}}^{\chi_{\rm max}^{\alpha}}\!\!\dd\chi\!\!\int_{\chi_{\rm min}^{\beta}}^{\chi_{\rm max}^{\beta}}\dd\chi'\int \dd k\,k^2 
(2/\pi) \frac{j_\ell(k\chi)}{a(\chi)^2} \frac{j_{\ell}(k\chi')}{a(\chi')^2}\sqrt{P^{\rm He}_{\Delta_e\Delta_e}(\chi)P^{\rm He}_{\Delta_e\Delta_e}(\chi')}\\
&+2\sigma_T^2 f_{\rm He} n_{\rm p,0}^2 \int_{\chi_{\rm min}^{\alpha}}^{\chi_{\rm max}^{\alpha}}\!\!\dd\chi\!\!\int_{\chi_{\rm min}^{\beta}}^{\chi_{\rm max}^{\beta}}\dd\chi'\int \dd k\,k^2 
(2/\pi) \frac{j_\ell(k\chi)}{a(\chi)^2} \frac{j_{\ell}(k\chi')}{a(\chi')^2}\\ 
&\hspace{18em} \times\sqrt{\bar{x}_e^{\rm He-H}(\chi)P(\chi)}\sqrt{\bar{x}_e^{\rm He-H}(\chi')P(\chi')}\,.
\end{split}
\ee
where $\bar{x}_e^{\rm He-H}(\chi)\equiv \bar{x}_e^{\rm H}(\chi)\bar{x}_e^{\rm He}(\chi)$ and the $k$-dependence of the power spectra ($P^{\theta}_{\Delta_e\Delta_e}$ and $P$) has been suppressed for ease of notation.
For the diagonal redshift bins, Eq.~\eqref{eq:cross_power_covariance} simplifies to
\begin{align}
    C_{\ell}^{\tau_\alpha \tau_\alpha}=\int_{\chi_{\rm min}^{\alpha}}^{\chi_{\rm max}^{\alpha}}\!\!\!\dd\chi\!\frac{\sigma_T^2n_{p,0}^2}{a^4\chi^2}P^{\rm H}_{\Delta_e\Delta_e}(\chi,k) &+ \int_{\chi_{\rm min}^{\alpha}}^{\chi_{\rm max}^{\alpha}}\!\!\!\dd\chi\!\frac{\sigma_T^2f_{\rm He}^2n_{p,0}^2}{a^4\chi^2}P^{\rm He}_{\Delta_e\Delta_e}(\chi,k) \nonumber \\ &+ 2\int_{\chi_{\rm min}^{\alpha}}^{\chi_{\rm max}^{\alpha}}\!\!\!\dd\chi \bar{x}_e^{\rm He}\bar{x}_e^{\rm H}\frac{\sigma_T^2f_{\rm He}n_{p,0}^2}{a^4\chi^2}P(\chi,k)\,,
    \label{eq:cl_tautauall_appendix}
\end{align}
where $k=\ell/\chi$, and we omitted showing the redshift dependence of mean ionization fractions for brevity. 

The cross power between the galaxy and the ionized electron fluctuations satisfies
\be
\begin{split}
C_{\ell}^{\tau_\alpha g_\beta}&=\sigma_T n_{\rm p,0} \int_{\chi_{\rm min}^{\alpha}}^{\chi_{\rm max}^{\alpha}}\!\!\dd\chi\!\!\int_{\chi_{\rm min}^{\beta}}^{\chi_{\rm max}^{\beta}}\dd\chi'\int \dd k\,k^2 
(2/\pi) \frac{j_\ell(k\chi)}{a(z)^2} {j_{\ell}(k\chi')}P_{\Delta x_e\,g}^{\rm H}(\chi,\chi',k)\\
&+\sigma_T f_{\rm He} n_{\rm p,0} \int_{\chi_{\rm min}^{\alpha}}^{\chi_{\rm max}^{\alpha}}\!\!\dd\chi\!\!\int_{\chi_{\rm min}^{\beta}}^{\chi_{\rm max}^{\beta}}\dd\chi'\int \dd k\,k^2 
(2/\pi) \frac{j_\ell(k\chi)}{a(z)^2} {j_{\ell}(k\chi')}P_{\Delta x_e\,g}^{\rm He}(\chi,\chi',k)\,,
\end{split}
\ee
where we defined
\be
P_{\Delta{x_e} g}(\chi,\chi',k)=[\bar{x}_e-(1-\bar{x}_e)\ln(1-\bar{x}_e )b\,I(k)]b_g\,\sqrt{P(\chi,k)}\sqrt{P(\chi',k)}\,.
\ee
Similarly for the diagonal redshift bins, the bin-bin auto-spectrum of the cross correlation reduces to the simpler form as follows:
\be
{C_{\ell}^{\tau_\alpha g_\alpha}=\int_{\chi_{\rm min}^{\alpha}}^{\chi_{\rm max}^{\alpha}}\!\!\!\dd\chi\!\frac{\sigma_Tn_{p,0}}{a^2\chi^2}P^{\rm H}_{\Delta_e g}(\chi,k)+\int_{\chi_{\rm min}^{\alpha}}^{\chi_{\rm max}^{\alpha}}\!\!\!\dd\chi\!\frac{\sigma_T f_{\rm He}n_{p,0}}{a^2\chi^2}P^{\rm He}_{\Delta_e g}(\chi,k)}\,,
\ee
with terms $P^{\rm H/He}_{\Delta x_e\,g}(\chi,k)$ defined in Eq.~\eqref{eq:powespec_cross}.

\section{Optical depth reconstruction noise}\label{sec:estimator_detail}

The modulations of temperature and polarization due to the effects described in Sec.~\ref{sec:effects_on_CMB} allows defining multiple estimators for the optical depth from a combination of temperature $T$ and polarization $E$ and $B$ fields. 
The corresponding $\Gamma$ couplings for these combinations are~\citep{Dvorkin:2008tf}
\be
\Gamma_{\ell_1\ell_2\ell}^{TT(\chi_\alpha)}=(C_{\ell_1}^{T_0T_1^\alpha}+C_{\ell_2}^{T_0T_1^\alpha})J_{000}^{\ell_1\ell_2\ell}\,,
\ee
\be
\Gamma_{\ell_1\ell_2\ell}^{EE(\chi_\alpha)}=\frac{1}{2}(C_{\ell_1}^{E_0E_1^\alpha}+C_{\ell_2}^{E_0E_1^\alpha})(J_{-220}^{\ell_1\ell_2\ell}+J_{2-20}^{\ell_1\ell_2\ell})\,,
\ee
\be
\begin{split}
\Gamma_{\ell_1\ell_2\ell}^{TE(\chi_\alpha)}=\frac{C_{\ell_1}^{T_0E_1^\alpha}}{2}(J_{-220}^{\ell_1\ell_2\ell}+J_{2-20}^{\ell_1\ell_2\ell})+C_\ell^{T_1E_0^\alpha}J_{000}^{\ell_1\ell_2\ell}\,,
\end{split}
\ee
\be
\Gamma_{\ell_1\ell_2\ell}^{TB(\chi_\alpha)}=\frac{C_{\ell_1}^{T_0E_1^\alpha}}{2i}(J_{-220}^{\ell_1\ell_2\ell}-J_{2-20}^{\ell_1\ell_2\ell})\,,
\ee
and 
\be
\Gamma_{\ell_1\ell_2\ell}^{EB(\chi_\alpha)}=\frac{C_{\ell_1}^{E_0E_1^\alpha}}{2i}(J_{-220}^{\ell_1\ell_2\ell}-J_{2-20}^{\ell_1\ell_2\ell})\,,
\ee
where 
\be
J_{m_1m_2m}^{\ell_1\ell_2\ell}=\sqrt{\frac{(2\ell_1+1)(2\ell_2+1)(2\ell+1)}{4\pi}}\begin{pmatrix}
\ell_1 & \ell_2 & \ell\\
m_1 & m_2 & m
\end{pmatrix}\,.
\ee
The (biased) noise terms can be found as 
\be
\begin{split}
\frac{1}{\tilde{N}_{\ell,\alpha\beta}^{\tau\tau;EB}}={\pi}\!\!\int\!\!\dd\theta\Big[\zeta^{E_0E_1(\chi_\alpha,\chi_\beta)}_{22}(\theta) \zeta^{B}_{22}-\zeta^{E_0E_1(\chi_\alpha,\chi_\beta)}_{2-2}(\theta) \zeta^{B}_{2-2}(\theta)\Big] d_{00}^{\ell}(\theta)\,,
\end{split}
\ee
where
\be
\zeta^{E_0E_1(\chi_\alpha)}_{mm'}(\theta)=\sum\limits_{\ell}\frac{(2\ell+1)}{4\pi}\frac{C_{\ell}^{E_0 E_1^\alpha}C_{\ell}^{E_0 E_1^\beta}}{C_{\ell}^{EE}+N_{\ell}^{EE}}d_{mm'}^{\ell}(\theta)\,,
\ee
\be
\zeta^{B}_{mm'}(\theta)=\sum\limits_{\ell}\frac{(2\ell+1)}{4\pi}\frac{1}{C_{\ell}^{BB}+N_{\ell}^{BB}}d_{mm'}^{\ell}(\theta)\,,
\ee
and
\be
\frac{1}{\tilde{N}_{\ell,\alpha\beta}^{\tau\tau;TT}}={4\pi}\!\!\int\!\!\dd\theta\! \left[\zeta^{T_0T_1(\chi_\alpha,\chi_\beta)}_{00}(\theta) \zeta^{T}_{00}+\zeta^{T_0T_1^\alpha}_{00,1}(\theta)\zeta^{T_0T_1^\beta}_{00,1}(\theta)\right] d_{00}^{\ell}(\theta)\,,
\ee
where
\be
\zeta^{T_0T_1(\chi_\alpha,\chi_\beta)}_{00}(\theta)=\sum\limits_{\ell}\frac{(2\ell+1)}{4\pi}\frac{C_{\ell}^{T_0 T_1^\alpha}C_{\ell}^{T_0 T_1^\beta}}{C_{\ell}^{TT}+N_{\ell}^{TT}}d_{00}^{\ell}(\theta)\,,
\ee
\be
\zeta^{T}_{00}(\theta)=\sum\limits_{\ell}\frac{(2\ell+1)}{4\pi}\frac{1}{C_{\ell}^{TT}+N_{\ell}^{TT}}d_{00}^{\ell}(\theta)\,,
\ee
\be
\zeta^{T_0T_1(\chi_\alpha)}_{00,1}(\theta)=\sum\limits_{\ell}\frac{(2\ell+1)}{4\pi}\frac{C_{\ell}^{T_0 T_1^\alpha}}{C_{\ell}^{TT}+N_{\ell}^{TT}}d_{00}^{\ell}(\theta)\,,
\ee
and
\be
\begin{split}
\frac{1}{\tilde{N}_{\ell,\alpha\alpha}^{\tau\tau;EE}}=2{\pi}\!\!\int\!\!\dd\theta \Big[&\zeta^{E_0E_1(\chi_\alpha,\chi_\beta)}_{22}(\theta) \zeta^{E}_{22} +\!\zeta^{E_0E_1(\chi_\alpha,\chi_\beta)}_{2-2}(\theta)\zeta^{E}_{2-2}(\theta) \\ &+\!\zeta^{E_0E_1(\chi_\alpha)}_{22,1}(\theta) \zeta^{E_0E_1(\chi_\beta)}_{22,1}(\theta)\! +\!\zeta^{E_0E_1(\chi_\alpha)}_{2-2,1}(\theta) \zeta^{E_0E_1(\chi_\beta)}_{2-2,1}(\theta)\Big] d_{00}^{\ell}(\theta)\,,
\end{split}
\ee
where
\be
\zeta^{E_0E_1(\chi_\alpha,\chi_\beta)}_{mm'}(\theta)=\sum\limits_{\ell}\frac{(2\ell+1)}{4\pi}\frac{C_{\ell}^{E_0 E_1^\alpha}C_{\ell}^{E_0 E_1^\beta}}{C_{\ell}^{EE}+N_{\ell}^{EE}}d_{mm'}^{\ell}(\theta)\,,
\ee
\be
\zeta^{E}_{mm'}(\theta)=\sum\limits_{\ell}\frac{(2\ell+1)}{4\pi}\frac{1}{C_{\ell}^{EE}+N_{\ell}^{EE}}d_{mm'}^{\ell}(\theta)\,,
\ee
\be
\zeta^{E_0E_1(\chi_\alpha)}_{mm',1}(\theta)=\sum\limits_{\ell}\frac{(2\ell+1)}{4\pi}\frac{C_{\ell}^{E_0 E_1^\alpha}}{C_{\ell}^{EE}+N_{\ell}^{EE}}d_{mm'}^{\ell}(\theta)\,,
\ee
and
\be
\frac{1}{\tilde{N}_{\ell,\alpha\beta}^{\tau\tau;TB}}={\pi}\!\!\int\!\!\dd\theta \left[\zeta^{T_0E_1(\chi_\alpha,\chi_\beta)}_{22}(\theta) \zeta^{B}_{22}{-}\zeta^{T_0E_1(\chi_\alpha,\chi_\beta)}_{2-2}(\theta) \zeta^{B}_{2-2}(\theta)\right] d_{00}^{\ell}(\theta)\,,
\ee
where
\be
\zeta^{T_0E_1(\chi_\alpha,\chi_\beta)}_{mm'}(\theta)=\sum\limits_{\ell_1}\frac{(2\ell_1+1)}{4\pi}\frac{C_{\ell_1}^{T_0 E_1^\alpha}C_{\ell_1}^{T_0 E_1^\beta}}{C_{\ell_1}^{TT}+N_{\ell_1}^{TT}}d_{mm'}^{\ell_1}(\theta)\,,
\ee
and finally
\be
\begin{split}
\frac{1}{\tilde{N}_{\ell,\alpha\beta}^{\tau\tau;TE}}={\pi}\!\!\int\!\!\dd\theta \big[&\zeta^{T_0E_1(\chi_\alpha,\chi_\beta)}_{22}(\theta) \zeta^{E}_{22}-\zeta^{T_0E_1(\chi_\alpha,\chi_\beta)}_{2-2}(\theta) \zeta^{E}_{2-2}(\theta)+\zeta^{T_0E_1(\chi_\alpha)}_{-20,1}(\theta)\zeta^{T_0E_1(\chi_\beta)}_{20,1}(\theta)\\
&\hspace{5em}+\zeta^{T_0E_1(\chi_\alpha)}_{20,1}(\theta)\zeta^{T_0E_1(\chi_\beta)}_{-20,1}(\theta)+2\zeta^{T_0E_1(\chi_\alpha,\chi_\beta)}_{00}(\theta)\zeta^{E}_{00}(\theta)\big] d_{00}^{\ell}(\theta)\,,
\end{split}
\ee
where
\be
\zeta^{T_0E_1(\chi_\alpha)}_{mm',1}(\theta)=\sum\limits_{\ell}\frac{(2\ell+1)}{4\pi}\frac{C_{\ell}^{T_0 E_1^\alpha}}{C_{\ell}^{TT}+N_{\ell}^{TT}}d_{mm'}^{\ell}(\theta)\,.
\ee
Throughout, we assume no tensor fluctuations and set $({\tilde{N}_{\ell,\alpha\beta}^{\tau\tau;BB}})^{-1}\rightarrow0$. These noise terms can be rotated using Eq.~\eqref{eq:rotation} to get the noise terms on the \textit{unbiased} optical-depth and the minimum-variance noise, which can be calculated using Eq.~\eqref{eq:minvar}.

\section{Improvements beyond ILC-cleaned foregrounds}\label{sec:imporvements_beyond_ilc}

\begin{figure}[h]
    \centering
    \includegraphics[width=0.9\textwidth]{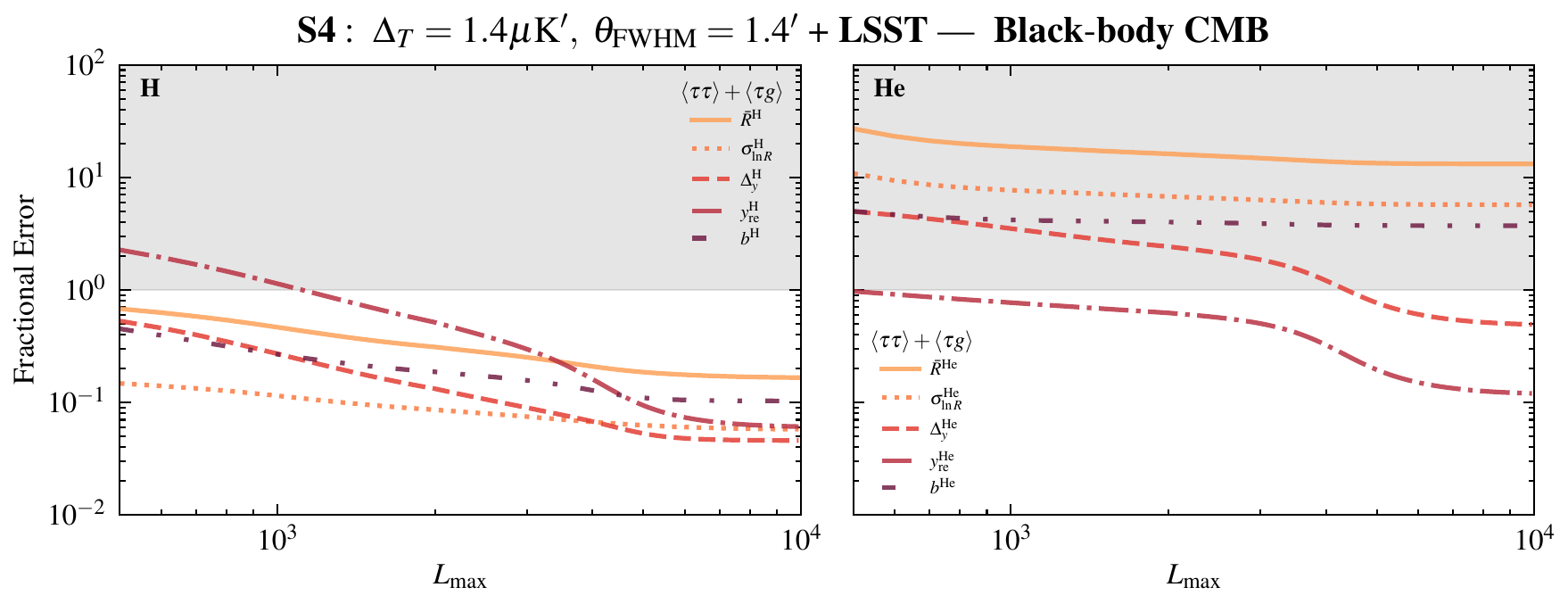} 
    \vspace*{-0.3cm}
    \caption{Similar to Fig.~\ref{fig:fisher_1D_ILC_S4}, but assuming black-body CMB spectra instead of ILC cleaning.}
    \label{fig:fisher_1D_BB_S4}
    \vspace*{0.5cm}
\end{figure}

\begin{figure}[h!]
    \centering
    \includegraphics[width=0.9\textwidth]{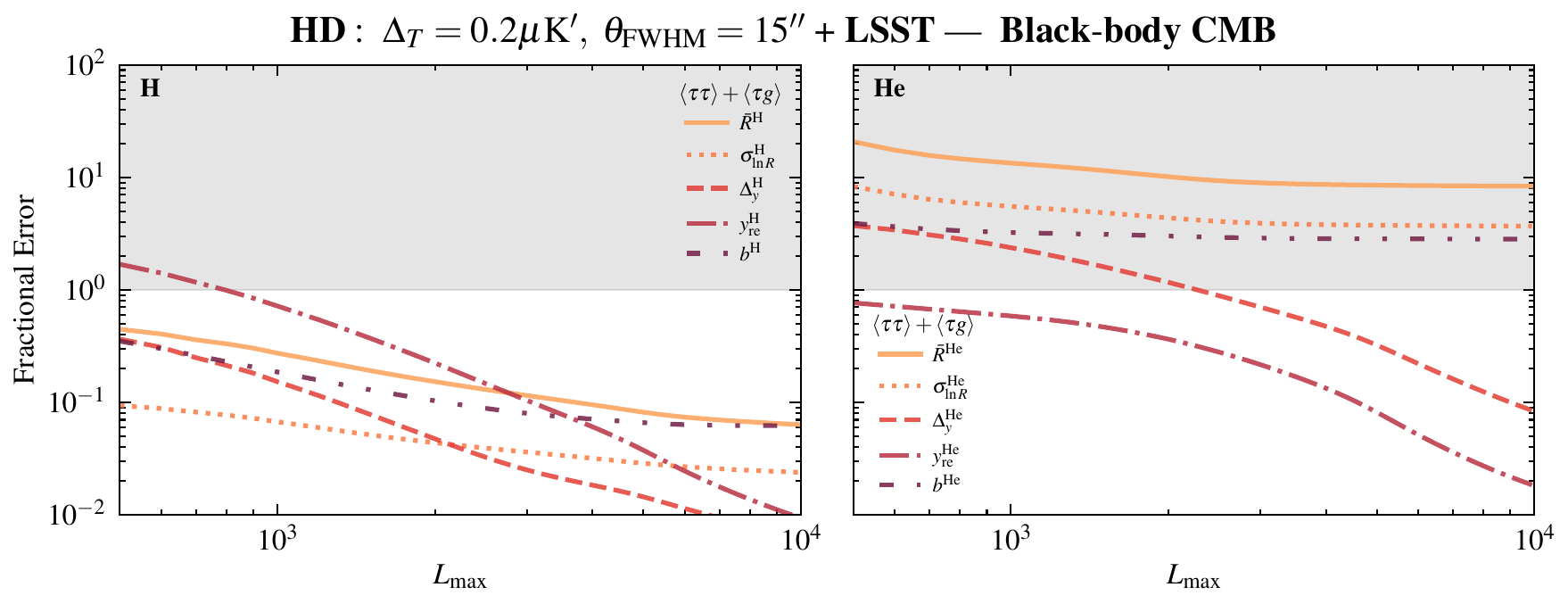} 
    \vspace*{-0.3cm}
    \caption{Similar to Fig.~\ref{fig:fisher_1D_ILC_HD}, but assuming black-body CMB spectra instead of ILC cleaning.}
    \label{fig:fisher_1D_BB_HD}
    \vspace*{0.5cm}
\end{figure}

\begin{figure}[h!]
    \centering
    \includegraphics[width=1\textwidth]{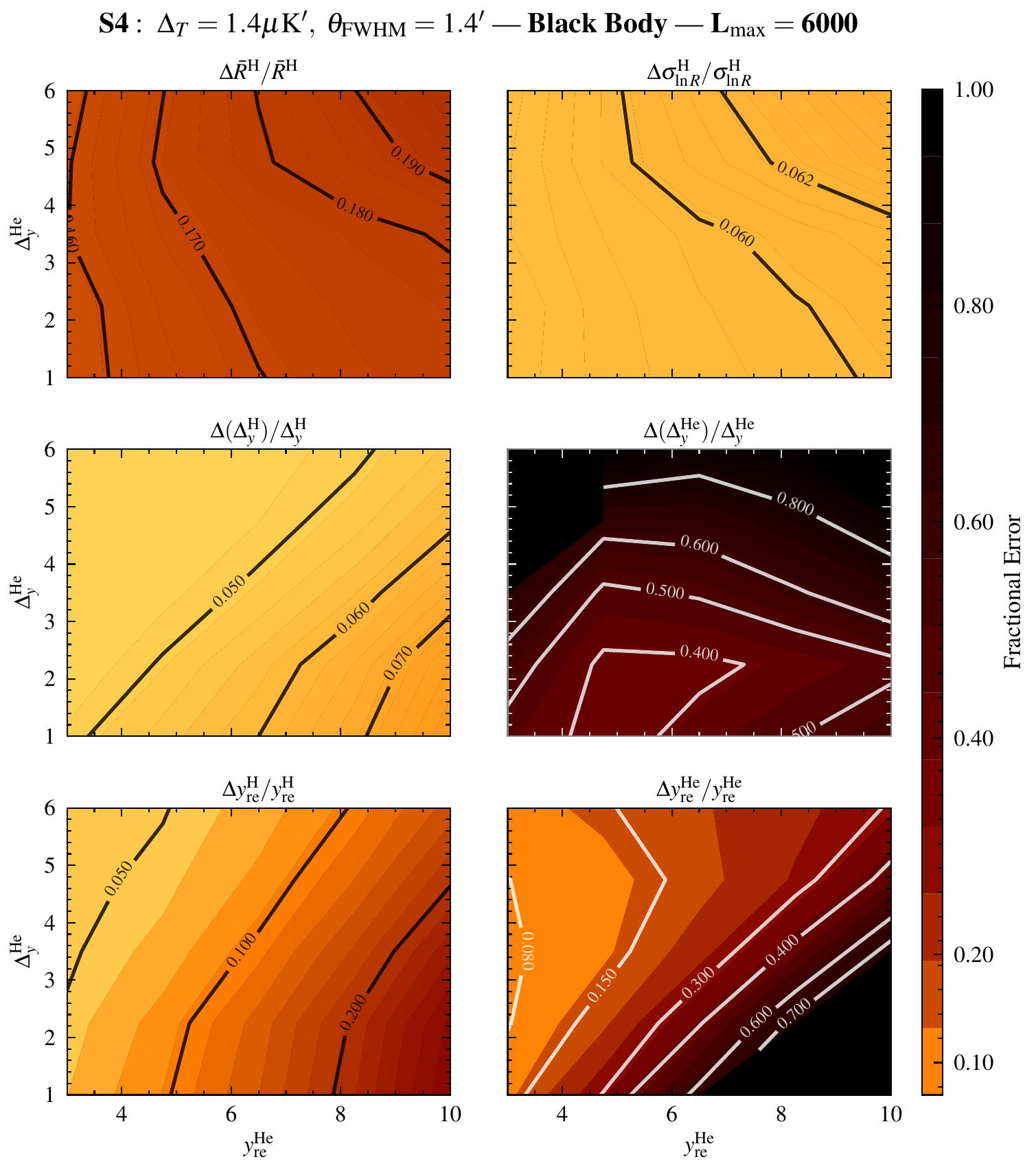} 
    \vspace*{-0.5cm}
    \caption{Similar to Fig.~\ref{fig:fisher_contour_ILC_S4}, but assuming black-body CMB spectra instead of ILC cleaning.
    }
    \label{fig:fisher_contour_BB_S4}
    \vspace*{-0.5cm}
\end{figure}

\begin{figure}[h!]
    \centering
    \includegraphics[width=1\textwidth]{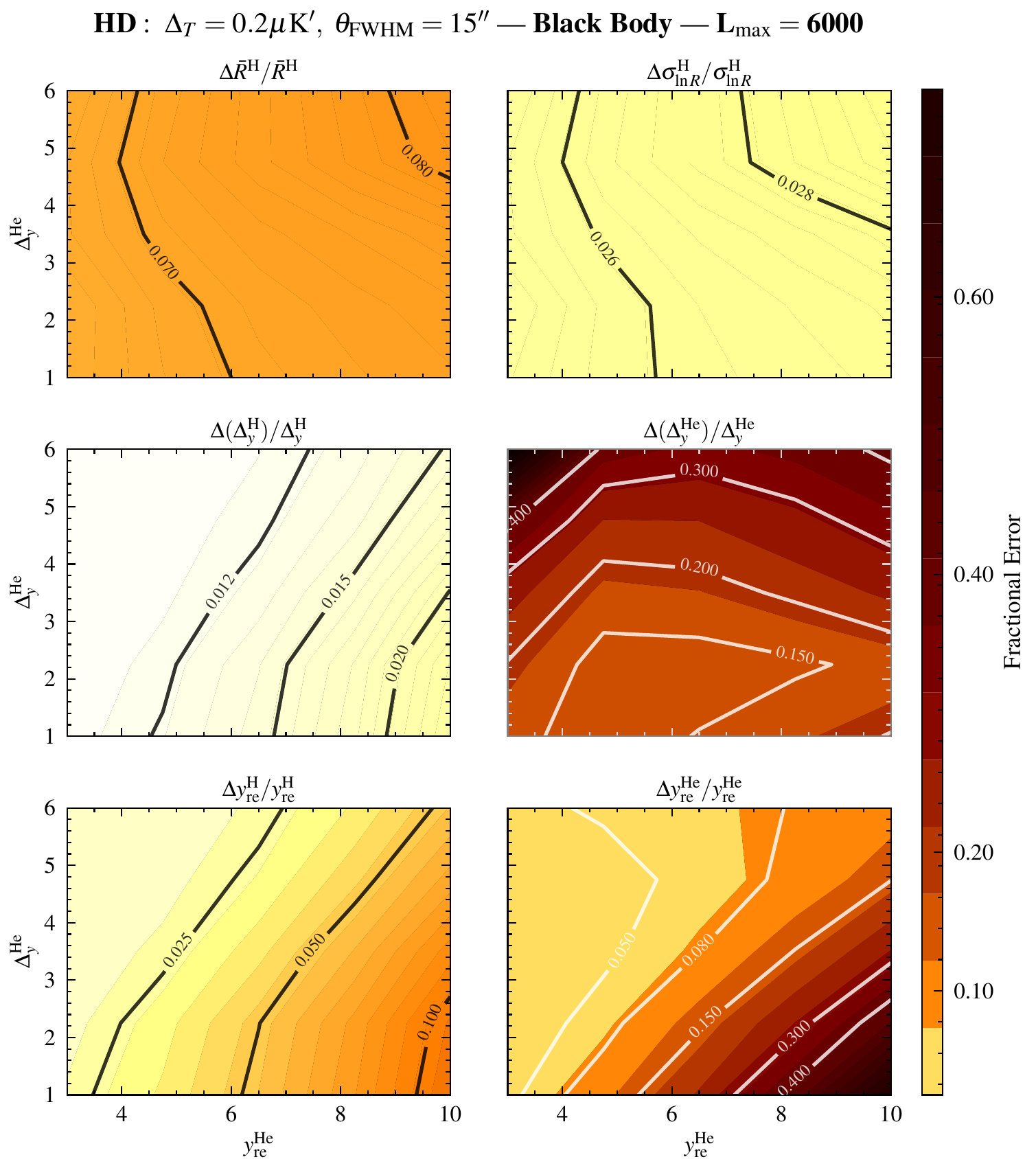} 
    \vspace*{-0.5cm}
    \caption{Similar to Fig.~\ref{fig:fisher_contour_ILC_HD}, but assuming black-body CMB spectra instead of ILC cleaning.}
    \label{fig:fisher_contour_BB_HD}
    \vspace*{-0.5cm}
\end{figure}

Our SNR and reionization parameter forecasts in Sec.~\ref{sec:forecasts} taken into account the residual foregrounds in the CMB maps after standard ILC cleaning. 
However, there is growing evidence that the foreground removal techniques will be improved in the near future~\citep[e.g.,][]{McCarthy:2023hpa,McCarthy:2023cwg,2009A&A...493..835D,2009ApJ...694..222C,2011MNRAS.410.2481R,2013MNRAS.435...18B} with more advanced methods like constrained-ILC~\citep{2011MNRAS.410.2481R} and needlet-based ILC cleaning~\citep{2013MNRAS.435...18B}. 
Better removal of frequency-dependent foregrounds improves the prospects of detecting signatures of helium reionization significantly, as can be seen from Figs.~\ref{fig:SNR_1} and~\ref{fig:SNR_2}. 
To highlight the reach of our method in case frequency-dependent foregrounds can be removed perfectly from the data, we show the SNR and parameter forecasts considering only the black-body contributions to the CMB in Figs.~\ref{fig:fisher_1D_BB_S4}--\ref{fig:fisher_contour_BB_HD}.

In Figs.~\ref{fig:fisher_1D_BB_S4}~and~\ref{fig:fisher_1D_BB_HD}, we show the fractional parameter measurement errors assuming black-body CMB spectra. 
The constraints on the helium reionization parameters are observed to improve. 
Most notably, our findings indicate that the helium reionization parameters, characterizing the time ($y_{\rm re}^{\rm He}$) and duration ($\Delta_y^{\rm He}$) of this epoch, become detectable with high significance for $L_{\rm max} > 3000$.

Figures~\ref{fig:fisher_contour_BB_S4} and~\ref{fig:fisher_contour_BB_HD} illustrate the fractional errors as functions of the varying values of $y_{\rm re}^{\rm He}$ and $\Delta_y^{\rm He}$, assuming black-body CMB-S4 and CMB-HD scenarios, respectively. 
We observe that the fractional errors improve by a factor of approximately 3 to 5 across all parameters. 
Compared to the ILC-cleaned CMB results shown in Figs.\ref{fig:fisher_1D_ILC_S4} and\ref{fig:fisher_1D_ILC_HD}, the dependency on the varying helium reionization parameter values is similar yet more pronounced.

\cleardoublepage

\bibliography{helium_reionization.bib}
\bibliographystyle{JHEP}

\end{document}